\pdfoutput=1
\UseRawInputEncoding
\documentclass[acmsmall,screen]{acmart}
\AtBeginDocument{%
  \providecommand\BibTeX{{%
    \normalfont B\kern-0.5em{\scshape i\kern-0.25em b}\kern-0.8em\TeX}}}

\setcopyright{acmcopyright}
\copyrightyear{2022}
\acmYear{2022}
\acmDOI{0000001.0000001}

\acmJournal{CSUR}
\acmVolume{36}
\acmNumber{4}
\acmArticle{66}
\acmMonth{7}

\usepackage{graphicx} 

\usepackage{amssymb}
\usepackage{bbm}
\usepackage{multirow}
\usepackage{longtable}
\usepackage{setspace}
\usepackage{url}

\usepackage{bm}
\usepackage{color}
\usepackage{amsthm}
\usepackage{subfigure}
\usepackage[figuresright]{rotating}
\usepackage{threeparttable}
\usepackage[T1]{fontenc}
\usepackage{hyperref}
\usepackage{setspace}

\usepackage{tikz}
\usetikzlibrary{positioning} %
\usepackage{xcolor} 
\usepackage{chapterbib}

\usepackage[utf8]{inputenc} 
\usepackage[T1]{fontenc}
\usepackage{amsthm}
\theoremstyle{definition}

\usepackage{titletoc}
\dottedcontents{section}[3.8em]{}{1.5em}{0.4pc}
\dottedcontents{subsection}[5.7em]{}{1.9em}{0.4pc}

\begin{document}


\newcommand{\bluepoint}{\textcolor{blue}{$\bullet$}} 
\newcommand{\cyanpoint}{\textcolor{cyan}{$\bullet$}} 
\newcommand{\graypoint}{\textcolor{gray}{$\bullet$}} 


\title{A Comprehensive Survey on Automatic Knowledge Graph Construction}

\author{Lingfeng Zhong}
\affiliation{%
 \institution{Macquarie University}
 \city{Sydney}
 \state{NSW}
 \country{Australia}}
\email{dmiczlf@gmail.com}

\author{Jia Wu}
\affiliation{%
 \institution{Macquarie University}
 \city{Sydney}
 \state{NSW}
 \country{Australia}}
\email{jia.wu@mq.edu.au}

\author{Qian Li}
\affiliation{%
 \institution{Beihang University}
 \city{Haidian}
 \state{Beijing}
 \country{China}}
\email{liqian@act.buaa.edu.cn}

\author{Hao Peng}
\affiliation{%
 \institution{Beihang University}
 \city{Haidian}
 \state{Beijing}
 \country{China}}
\email{penghao@act.buaa.edu.cn}

\author{Xindong Wu}
\affiliation{%
 \institution{Hefei University of Technology}
 \city{Hefei}
 \country{China}}
\email{uvmxwu@gmail.com}

\renewcommand{\shortauthors}{Lingfeng Zhong, et al.}

\begin{abstract}

Automatic knowledge graph construction aims to manufacture structured human knowledge. To this end, much effort has historically been spent extracting informative fact patterns from different data sources. However, more recently, research interest has shifted to acquiring conceptualized structured knowledge beyond informative data. In addition, researchers have also been exploring new ways of handling sophisticated construction tasks in diversified scenarios. Thus, there is a demand for a systematic review of paradigms to organize knowledge structures beyond data-level mentions. To meet this demand, we comprehensively survey more than 300 methods to summarize the latest developments in knowledge graph construction. A knowledge graph is built in three steps: knowledge acquisition, knowledge refinement, and knowledge evolution. The processes of knowledge acquisition are reviewed in detail, including obtaining entities with fine-grained types and their conceptual linkages to knowledge graphs; resolving coreferences; and extracting entity relationships in complex scenarios. The survey covers models for knowledge refinement, including knowledge graph completion, and knowledge fusion. Methods to handle knowledge evolution are also systematically presented, including condition knowledge acquisition, condition knowledge graph completion, and knowledge dynamic. We present the paradigms to compare the distinction among these methods along the axis of the data environment, motivation, and architecture. Additionally, we also provide briefs on accessible resources that can help readers to develop practical knowledge graph systems. The survey concludes with discussions on the challenges and possible directions for future exploration.

\end{abstract}



\maketitle
\section{Introduction}

Knowledge graphs (KGs) provide well-organized human knowledge for applications like search engines \cite{DBLP:conf/sigir/0006RZCL020}, recommendation systems \cite{DBLP:conf/kdd/Wang00LC19}, and question answering \cite{DBLP:conf/coling/BaoDYZZ16}.

Many of the well-known large KG systems have been constructed through crowd-sourcing, like Freebase \cite{DBLP:conf/aaai/BollackerCT07} and Wikidata \cite{DBLP:conf/www/Vrandecic12}. Hence, a systematic solution that can automatically build a knowledge graph from unstructured or semi-structured data offers a massive boost to what is a very arduous manual process for practical purposes.

\begin{figure*}[!t]
 \centering

 \includegraphics[width=\linewidth]{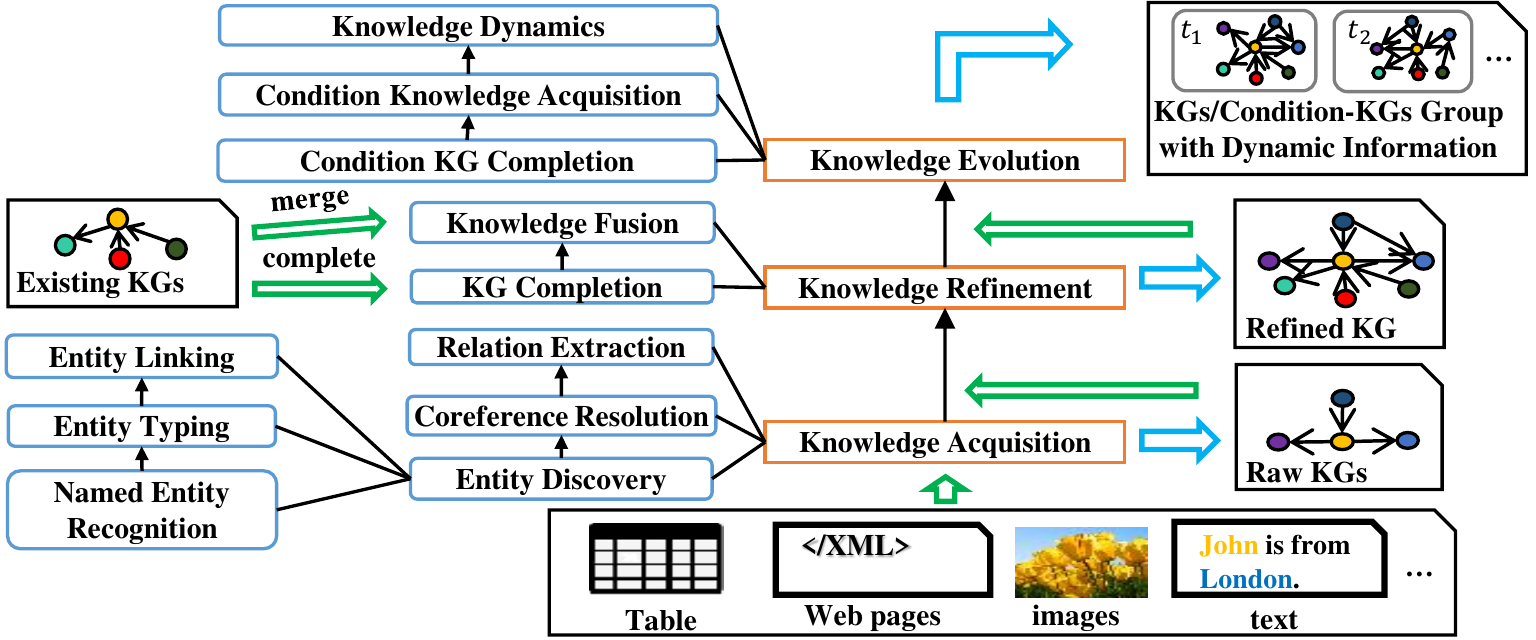}
 \caption{The general process of constructing a knowledge graph. In this diagram, semi-structured or unstructured input data is manufactured into a raw knowledge graph by acquiring knowledge. Then the knowledge will be refined to complete the knowledge graph or enrich it with other existing knowledge graphs. If the input is only an existing knowledge graph, it will be directly handled by the knowledge refinement process. Last, the knowledge evolution process will try to obtain a group of knowledge graphs/conditional knowledge graphs that contains dynamic information about the graph’s evolution.}
 \label{fig:figure11}
 \vspace{-0.3cm}
\end{figure*} 

A knowledge graph is a semantic graph consisting of edges and nodes that depicts knowledge of real-world objects. Within these structures, a knowledge tuple is the minimum knowledge-carrying group. The tuples comprise two nodes representing concepts connected by an edge representing a relationship. Thus, constructing a knowledge graph is the task of discovering the elements that constitute a knowledge graph in a domain-specific or open-domain area. Early in the study of this discipline, researchers were mostly focused on scratching out factual tuples from semi-structured or unstructured textual data as patterned knowledge mentions.  Information extraction systems like TextRunner  \cite{DBLP:conf/naacl/YatesBBCES07} and  Knowitall \cite{DBLP:conf/www/EtzioniCDKPSSWY04} are the milestones for early knowledge graph construction, driven by designated rules or clustering. Unfortunately, these designs are not sufficiently equipped with background knowledge, and thus suffer from two major defects: 1) insubstantial, traditional information extraction systems do not create or distinguish entities from different expressions, which prevents knowledge aggregation; 2) uninformative, traditional information extraction systems only extract information from syntactic structures without capturing the semantic denotations in the given expressions. Furthermore, conventional rule-based information extraction systems also require heavy feature engineering and extra expert knowledge. Wu et al. \cite{DBLP:journals/kais/WuW19} point out that if a KG system does not organize nodes and edges with background knowledge about 
concepts, it is merely a data graph.

Regarding this issue, researchers then recourse to well-partitioned acquisition sub-tasks for arranging semantic knowledge structures. The most classic paradigm is the pipeline that first discovers and links conceptual entities, resolves coreference mentions, then extracts relationships among entities. The general procedure of knowledge graph  construction is displayed in Fig. \ref{fig:figure11}. 

More recently, deep learning methods have given rise to tremendous breakthroughs in natural language processing (NLP), and these breakthroughs have fed applications for knowledge graph construction in a range of respects. Numerous deep learning models have delivered good performances with tasks like named entity recognition \cite{DBLP:journals/corr/HuangXY15}\cite{DBLP:conf/acl/MaH16}, entity typing  \cite{DBLP:conf/naacl/XuB18}\cite{DBLP:conf/kdd/RenHQVJH16}, entity linking  \cite{DBLP:conf/emnlp/GaneaH17}\cite{DBLP:conf/acl/TitovL18a}, coreference resolution  \cite{DBLP:conf/emnlp/LeeHLZ17}, relation extraction \cite{DBLP:conf/emnlp/ZengLC015}\cite{DBLP:conf/acl/ZhouSTQLHX16}. Additionally, deep knowledge representation models have also been developed that can refine knowledge graphs. The refinements include completing corrupt tuples, discovering new tuples in a built knowledge graph via its inner graph structure, and merging graphs from different sources to construct new knowledge graphs. At present, many knowledge bases\footnote{Knowledge base (KB) and knowledge graph (KG) are identical terms in this paper.}, such as TransOMCS \cite{DBLP:conf/ijcai/ZhangKSR20}, ASER \cite{DBLP:journals/corr/abs-2104-02137} and huapu \cite{xin2020huapu} have put automatic KG construction methods into practice.

Further, with advances in the pre-training of deep learning models, such as the pre-trained BERT model \cite{DBLP:conf/naacl/DevlinCLT19} and some of the massive-scale graph convolution network (GCN) models, KG construction tasks are being applied to more complicated scenarios in the big data environment. Beyond systems that deal with heterogeneous data, like web pages and table forms, more attention is being paid to effective methods of tackling complex data – for example, jointly unifying multiple acquisition sub-tasks or solutions that harvest knowledge graphs from long-term contexts \cite{DBLP:conf/acl/YaoYLHLLLHZS19}, noisy data \cite{DBLP:conf/pkdd/RiedelYM10}, or low-resource data \cite{DBLP:conf/coling/WangHL0S18}.

\begin{figure*}[!t]
 \centering
 \scriptsize
 \includegraphics[width=\linewidth]{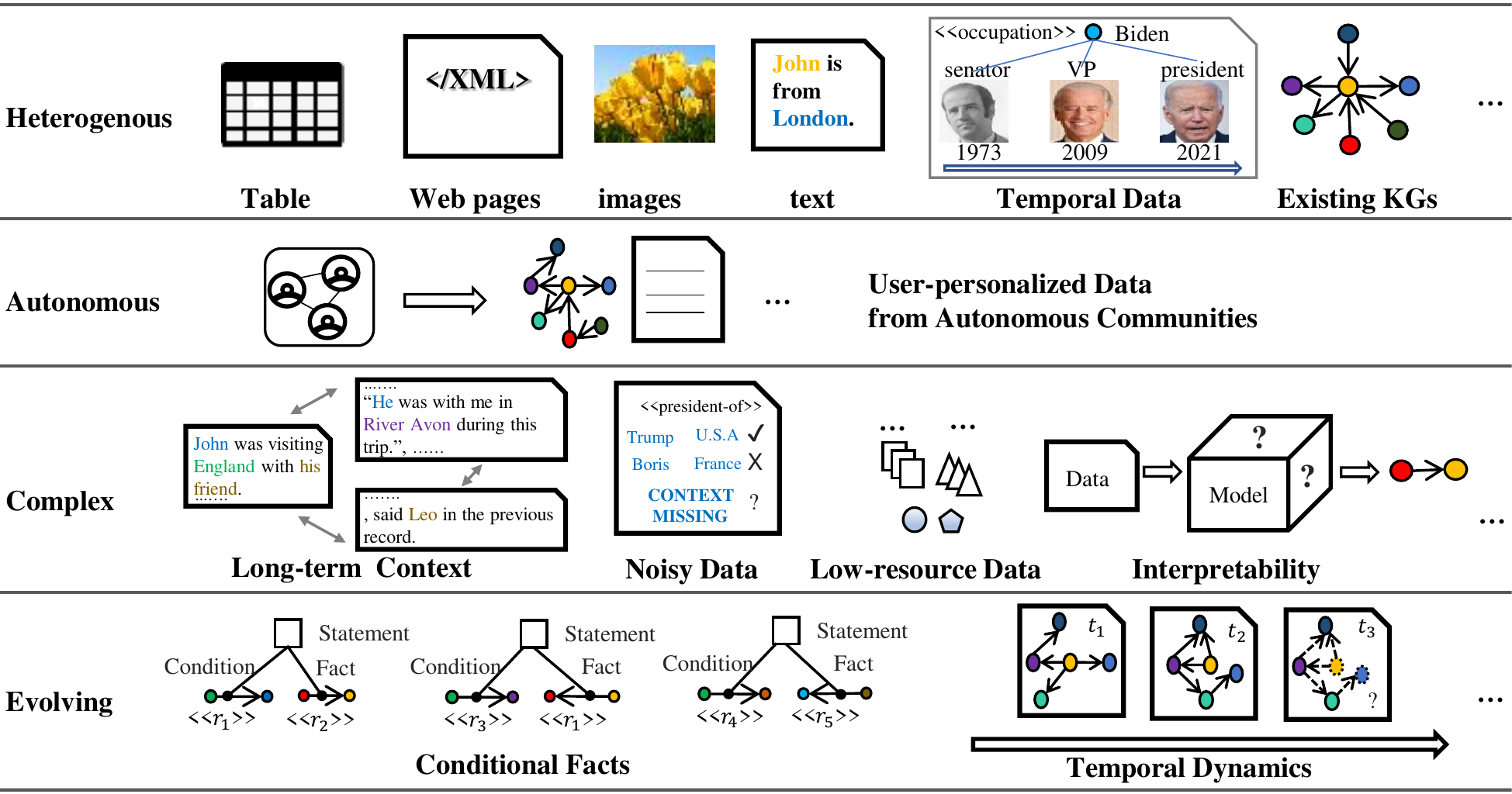}
 \caption{An illustration of the challenges framed by the HACE environments. In terms of heterogeneous data, knowledge graph construction with semi-structured and unstructured data in outlined in Section \ref{Section 4}. Methods of refining existing knowledge graphs are detailed in Section \ref{Section 5}. Methods of obtaining temporal data are described in \ref{cka}. Section \ref{crossmodal} presents a discussion on multi-modal knowledge graphs. In terms of complex data, long-term contexts and their involvement with multiple KG construction tasks are discussed in Sections  \ref{et}, \ref{CO}, \ref{docre}. Methods for tackling noisy data are mainly presented in Section \ref{dsre}. Model interpretability is covered in Section  \ref{interp}. In terms of evolving data, recent work in knowledge evolution is presented in Section  \ref{Section 7}, and research on autonomous data is discussed in Section \ref{AutonomousKGC}.}
 
 \label{fig:figure12}
 \vspace{-0.38cm}
\end{figure*} 

In terms of knowledge graph refinement tasks, interpretable reasoning has become a prevalent trend. Researchers are seeking solutions that merge cross-lingual knowledge and derive new relationships between nodes through logic and reasoning. Researchers are also focusing on knowledge graphs for conditional knowledge, such as temporal knowledge graphs \cite{DBLP:conf/iclr/HanCMT21} and the generic condition knowledge graphs  \cite{DBLP:conf/emnlp/JiangZQLCJ19}. Active learning \cite{DBLP:conf/ijcnn/PradhanTSS20}, which asks human users about unknown valuable data for collection, is another significant direction for handling knowledge from autonomous communities. Wu et al. \cite{DBLP:journals/tkde/WuZW014} summarize the challenges facing knowledge discovery in big data environments with the HACE theorem, which is shown in Fig. \ref{fig:figure12}.

\subsection{Major Differences and Contributions}

\begin{table}[!t]
\centering
\caption{
A Comparison between Existing Survey on Knowledge Graph Construction. 
}
\label{tab1}

\resizebox{\textwidth}{!}{

\begin{tabular}{l|l|c|c|c|c|c|c|c|c|c|c|c|c|c|c}
\toprule[1.5pt]
\multirow{2}{*}{Survey} & \multirow{2}{*}{Year} & \multirow{2}{*}{Topic} & \multicolumn{4}{c}{KA} &
\multicolumn{3}{|c|}{KGR} &
\multicolumn{4}{c}{Target Data} & 
\multicolumn{2}{|c}{Resource}
\\ 
\cline{4-16}
 & & & Ent & Rel & CO & Cond & KGC & TKGC & KF & 
 Web & Tab & Sent & Doc & Tool & Dataset \\
\midrule[1.5pt]
\textbf{Our Survey} & 2022 & Overall Process of KG Construction in HACE environment& \bluepoint & \bluepoint & \bluepoint & \cyanpoint & \bluepoint & \bluepoint & \bluepoint & \bluepoint & \bluepoint & \bluepoint & \bluepoint & \bluepoint & \bluepoint \\

Paulheim \cite{DBLP:journals/semweb/Paulheim17} & 2017 & KG Refinement & \graypoint & \cyanpoint & - & - & \bluepoint & - & - & \graypoint & \graypoint & \cyanpoint & - & - & \cyanpoint \\
Yan et al. \cite{DBLP:journals/fcsc/YanWCGZ18} & 2018 & KG Application, Construction & \bluepoint & \bluepoint & \graypoint & - & \bluepoint & - & - & \cyanpoint & \graypoint & \bluepoint & \graypoint & \bluepoint & \bluepoint \\

Wu et al. \cite{DBLP:conf/icdm/0001WFLZJ19} & 2019 & Raw KG Construction & \cyanpoint & \cyanpoint & \cyanpoint & - & - & - & - & \graypoint & - & \bluepoint & - & \cyanpoint & \graypoint \\

Ji et al. \cite{DBLP:journals/corr/abs-2002-00388} & 2020 & KG Application, Representation, Acquisition & \bluepoint & \bluepoint & - & - & \bluepoint & \bluepoint & \cyanpoint & - & - & \bluepoint & - & \cyanpoint & \bluepoint \\

Arora \cite{DBLP:journals/corr/abs-2007-12374} & 2020 & KG Completion & - & - & - & - & \bluepoint & - & - & - & - & - & - & - & \graypoint \\
Nayak et al. \cite{DBLP:journals/cogcom/NayakMGP21} & 2021 & Relation Triples Extraction & \bluepoint & \bluepoint & - & - & - & - & - & \graypoint & - & \bluepoint & \graypoint & \graypoint & \cyanpoint \\
Pawar et al. \cite{DBLP:journals/corr/abs-2103-06118} & 2021 & Joint Extraction of Entities and Relations & \bluepoint & \bluepoint & - & - & - & - & - & - & - & \bluepoint & - & - & \graypoint \\
Hogan et al. \cite{DBLP:series/synthesis/2021Hogan} & 2021 & Overview of KG & \bluepoint & \bluepoint & - & - & \bluepoint & \graypoint & - & \cyanpoint & \bluepoint & \bluepoint & - & \cyanpoint & \bluepoint \\

Cai et al. \cite{DBLP:journals/corr/abs-2201-08236} & 2022 & Temporal KG & - & - & - & - & \bluepoint & \bluepoint & - & - & - & - & - & - & \graypoint \\
\bottomrule[1.5pt]
\end{tabular}
}
\flushleft
{
*-:Not covered, \graypoint:1-5 references covered, \cyanpoint:6-14 references covered, \bluepoint: 15+ references covered. \\
*Ent: Entity, 
Rel: Relationship, 
CO: Coreference, 
Cond: Condition (timestamp or prerequisite). 
*Web: Web page, Tab: Table forms, Sent: Sentence, Doc: Textual document.\\
*KA: Knowledge Acquisition, KGR: Knowledge Graph Refinement, KF: Knowledge Fusion.\\
*KGC: Knowledge Graph Completion,  TKGC: Temporal Knowledge Graph Completion.
}
\vspace{-0.3cm}
\end{table}

Many surveys have provided an overview of knowledge graphs and their applications. For example, Hogan et al. \cite{DBLP:series/synthesis/2021Hogan} provided an encyclopedic survey for the knowledge graph, while Paulheim \cite{DBLP:journals/semweb/Paulheim17} looks into methods that refine and fill knowledge graphs. Other surveys summarize methods for acquiring knowledge from unstructured or semi-structured data. Wu et al. \cite{DBLP:conf/icdm/0001WFLZJ19} review competitive tools and models for KG construction sub-task over texts including relation extraction, named entity recognition, and coreference resolution, while Yan et. al \cite{DBLP:journals/fcsc/YanWCGZ18} browse methods for different data types like web pages, table forms, etc. Deep learning approaches for jointly extracting entities with their relationships are reviewed in \cite{DBLP:journals/cogcom/NayakMGP21}, \cite{DBLP:journals/corr/abs-2103-06118}. Some surveys also focus on acquiring knowledge from existing knowledge graphs. Prior work such as \cite{DBLP:journals/corr/abs-2002-00388} and \cite{DBLP:journals/corr/abs-2007-12374} also covers the methods for knowledge representation learning and knowledge graph completion, while Cai et al. \cite{DBLP:journals/corr/abs-2201-08236} dive into temporal knowledge graphs. Table \ref{tab1} compares previous work with this survey. 

Unlike other surveys, we go deeper into the paradigms of the recent models for knowledge graph construction, arranging our work according to different stages and aspects of the HACE environments. We also present practical resources and discuss future challenges and directions with data, models, and architectures. Hence, our contributions are summarized as follows:

\begin{itemize}
 \item We introduce the process of knowledge graph construction and various knowledge graphs by giving formal definitions and classifications. We also summarize necessary information on KG-related resources, including practical knowledge graph projects and construction tools, covering published years, citations, and access links for readers to compare.

\item We comprehensively analyze models for knowledge graph construction in different scenarios – from knowledge acquisition to knowledge graph refinement – according to their task backgrounds and challenges. We summarize the motivations and designs of classical and novel models, then primarily delineate the pragmatics in terms of their architectures and improvements.

 \item We discuss knowledge graph construction in HACE big data environments, including noisy, document-level data and low-resource data. Then we review achievements for acquiring model interpretability and evolutionary condition knowledge. Finally, we summarize the major challenges and directions that impact KG construction tasks.

\end{itemize}

\begin{table}[!t]
\centering

\caption{The information of practical KG projects.}
	\resizebox{\textwidth}{!}{
\begin{tabular}{lllll}
\toprule[1.75pt]
Categorization&Project& KG Inclusion& Year& URL\\

\midrule[1.75pt] 

\multirow{7}{*}{Encyclopedia KG} &YAGO & 2B+ facts, 64M+ entities & 2007 & https://yago-knowledge.org\\
&Freebase & 360M+ fact triples & 2007 & https://freebase-easy.cs.uni-freiburg.de/dump/\\
&DBpedia & 320 classes with 1, 650 different properties, 247M+ triples & 2007 & https://github.com/DBpedia/\\
&CN-DBpedia & 9M+ entities, 67M+ triples & 2015 & http://kw.fudan.edu.cn/cnDBpedia/download/\\ 
&Probase & 5.4M+ concepts & 2010 & https://concept.research.microsoft.com/\\
&Wikidata & 96M+items & 2012 & https://www.wikidata.org/wiki\\
&CN-Probase & 17M+ entities, 33M+ ``is-a'' relations & 2017 & http://kw.fudan.edu.cn/apis/cnprobase/\\
\hline
\multirow{5}{*}{Linguistic KG}&WordNet & 117 000 synsets & 1985 & https://wordnet.princeton.edu/\\
&ConceptNet & 34M+ items & 1999 & https://www.conceptnet.io/\\
&HowNet & 35, 202 concepts, 2, 196 sememes & 1999 & https://openhownet.thunlp.org/download\\
&Babelnet & 13M nodes & 2010 & http://babelnet.org/rdf/page/\\
&THUOCL & 157K+ word nodes in 7.3B+ documents & 2016 & http://thuocl.thunlp.org/\\
\hline
 
\multirow{3}{*}{Commonsense KG}&OpenCyc & 2M+ fact triples & 1984 & https://sourceforge.net/projects/opencyc/\\
&ASER & 438M+ nodes, 648M+ edges & 2020 & https://github.com/HKUST-KnowComp/ASER\\
&TransOMCS & 18M+ + & 2020 & https://github.com/HKUST-KnowComp/TransOMCS\\ 
\hline

\multirow{2}{*}{Enterprise support KG}&Google Knowledge Graph & 500B+ facts on 5B+ entities & 2012 & https://developers.google.com/knowledge-graph\\
&Facebook Graph Search & dynamic social network of users, User-generated contents & 2013 & https://developers.facebook.com/docs/graph-api/\\
\hline
\multirow{8}{*}{Domain-specific KG} &Pubmed & 5B+triples & 2000 & http://pubmed.bio2rdf.org/sparql\\

&Drugbank & 14K+ drug entities & 2006 & https://go.drugbank.com/releases/latest\\
&AMiner ASN & 2M+ paper nodes, 8M+ citation relations & 2007 & https://www.aminer.cn/aminernetwork\\
&Huapu & 17M+ person nodes & 2017 & https://www.zhonghuapu.com/\\
&OAG & 369M+ authors, 380M+ papers 92M+ linking relations & 2017 & https://www.aminer.cn/data/?nav=openData\#Open-Academic-Graph\\
&TransOMCS & 18M+ + & 2020 & https://github.com/HKUST-KnowComp/TransOMCS\\ 

&COVID-19 Concepts & 4784 entities, 35172 relation links & 2020 & http://openkg.cn/dataset/covid-19-concept\\

&Aminer COVID-19 Open Data & reports, news, research and other achieves of COVID-19 & 2020 & https://www.aminer.cn/data-covid19/\\
\hline
\multirow{2}{*}{Federated KG} 

&GEDMatch & 1.2M+ DNA profiles & 2010 & https://www.gedmatch.com/\\
&OpenKG.cn & 200+ datasets from 94 organizations & 2015 & http://www.openkg.cn\\

\bottomrule[1.75pt] 
\end{tabular}
\vspace{-0.3cm}
}
\label{tabKGPrj}
\end{table}
\subsection{Organization of the Survey}
We organize our survey as follows. 
Section~\ref{Section 2} gives the background for knowledge graph construction, including definitions and resources of KG projects. Section~\ref{Section 3} delivers methods to pre-process semi-structured data, including content extraction and structure interpretation. 
Section~\ref{Section 4} introduces the methods for handling tasks that obtain entities and relationships from various data types and environments.
Section~\ref{Section 5} reviews the methods for refining knowledge graphs with external structured data.  
 We portray recent achievements and trends in evolutionary knowledge graphs, including conditional knowledge graphs and temporal knowledge graphs in Section ~\ref{Section 7}.  Section~\ref{Section KGStorage} delivers solutions that store knowledge graphs effectively.
Finally, we envisage future directions and development in Section ~\ref{Section 8} while concluding the article in Section~\ref{Section 9}.

\section{Background}
\label{Section 2}
\subsection{Definitions}
Many contributions have been made to formally define knowledge graphs. Wang et al. \cite{DBLP:journals/tkde/WangMWG17} modeled a knowledge graph as a multi-relation graph, where the nodes are entities and the edges represent different types of relationships. However, the previous definition does not consider the semantic structures in a knowledge graph. Ehrlinger and Woß \cite{DBLP:conf/i-semantics/EhrlingerW16} further emphasize that a knowledge graph arranges information into an ontology and then enlightens novel knowledge discovery with ``a reasoner''. To specifically protrude the essence component supporting knowledge-level information, Wu et al. \cite{DBLP:conf/icdm/0001WFLZJ19} define a knowledge graph as a semantic graph where the nodes represent concepts (entities/attributes/facts), and the edges represent relationships that connect the nodes while drawing on background knowledge about the concepts and relations. 
\paragraph{\textbf{Definition 1 (Knowledge Graph)}} A knowledge graph $\mathcal{G}$ is defined as $\mathcal{G} = \{\mathcal{E}, \mathcal{R}, \mathcal{T}, \mathcal{F}_k \} $, where $\mathcal{E}$ and $\mathcal{R}$ represent sets of concepts and relations, respectively. In this paper, concepts can be regarded as entities/attributes. $\mathcal{T}$ is the set of factual triples, where a standard binary fact is a triple $ (h, t, r) \in \mathcal{T}, h, t \in \mathcal{E}, r \in \mathcal{R} $. An n-ary relation triple will be formed as $ (e_1, ..., e_n, r)$, where $e_1, ..., e_n \in \mathcal{E}$. $\mathcal{F}_k$ is a set function representing the background knowledge that constrains potential facts to be knowledge-level informative, and we have $ \mathcal{T} \subset \mathcal{F}_k (\{\mathcal{E}, \mathcal{R} \} ) $.  In practice, background knowledge can be seen as a rule set, schema, or set of implicit math principles.

\paragraph{\textbf{Definition 2 (Knowledge Graph Construction)}} Knowledge graph construction $f$ is a procedure that maps a data source into a knowledge graph: $f: D \times f_{k} (D) \rightarrow \mathcal{G} $, where D is the set of data sources, and $f_{k} (D)$ is background knowledge of the data target, which can be domain knowledge. Notably, knowledge graph construction is usually unable to continue without background knowledge that is provided by pre-designed rules or a language model of representations.

\subsection{Practical Knowledge Graph Projects}
In this section, we review the representative practical projects (datasets) of knowledge graphs, including encyclopedia knowledge graphs, linguistic knowledge graphs, commonsense knowledge graphs, enterprise support knowledge graphs, domain-specific knowledge graphs, and federated knowledge graphs. The details are presented in Table \ref{tabKGPrj}.

\subsubsection{Encyclopedia KGs}
\
\par
 Encyclopedia knowledge graphs systematically cover factual or event knowledge from different domains. Many researchers have developed knowledge graph structures from manually-built online encyclopedias. For example, DBpedia \cite{DBLP:conf/semweb/AuerBKLCI07} (developed from Wikipedia) is a fundamental encyclopedia knowledge graph, while Freebase \cite{DBLP:conf/aaai/BollackerCT07} incorporates automatic extraction tools to obtain more content. Probase \cite{DBLP:conf/sigmod/WuLWZ12} (supported by Microsoft Concept Graph), as an event encyclopedia knowledge graph, creatively depicts knowledge of uncertain events containing conflicting information in the form of probabilistic models. XLore \cite{DBLP:conf/semweb/WangLWLLZSLZT13} (a sub-project of THUKC), as a multi-lingual encyclopedia knowledge graph, establishes entity links across multi-lingual content via deep learning approaches. After an early attempt by the DBpedia project to incorporate automatic extraction tools, more knowledge graph projects decided to do the same, including Wikidata \cite{DBLP:conf/www/Vrandecic12} and CN-DBpedia \cite{DBLP:conf/ieaaie/XuXLXLCX17}. The Max Planck Institution developed YAGO \cite{DBLP:conf/www/SuchanekKW07}, which integrates temporal and geographical structures in Wikipedia with a WordNet ontology. Minz et al. \cite{DBLP:conf/acl/MintzBSJ09} applied distance supervision to Freebase for automatically annotating entity relationships. The research community has also been concerned with knowledge graphs of eventualities. For example, CN-Probase \cite{DBLP:conf/icde/ChenWCXCLLW19} extends Probase with concepts in Chinese to understand general modes of textual data that involve uncertain occurrences.

\subsubsection{Linguistic KGs}
\
\par

Linguistic knowledge graphs deliver knowledge of the human language to provide basic semantics as ontologies or external features. WordNet \cite{DBLP:journals/cacm/Miller95} is a classical widely-used knowledge graph dictionary for linguistic study, providing synonymy or hyponymy relationships among words. With these tools, developers create high-performance word embeddings based on well-built linguistic knowledge graphs for downstream applications. Beyond WordNet , BabelNet \cite{DBLP:conf/acl/NavigliP10} extends WordNet with cross-lingual attributes and relations of words from encyclopedias. ConceptNet \cite{liu2004conceptnet}, as a part of Link Open Data, gathers conceptual knowledge based on crown sourcing, while HowNet \cite{1276017} manually collects sememe information (minimum indivisible semantic units)  about word concepts and attributes. THUOCL \cite{han2016thuocl} records the document frequency of words from a well-filtered web corpus. Developers have also created high-performance word embeddings based on well-built linguistic knowledge graphs for downstream applications.

\subsubsection{Enterprise support KGs}
\
\par

Knowledge graphs and their related systems have been effectively supporting the business of enterprise activities. Google Knowledge Graph (GKG) \cite{DBLP:conf/semweb/SteinerVTGW12}, served as a core function since 2012, delivers knowledge support for user queries and enriches the results with more semantically-related content. Facebook Graph Search \footnote{https://developers.facebook.com/docs/graph-api/} is the powerful semantic search engine of Facebook, providing user-specific answers through the dynamic knowledge base in the Facebook social system. Similar to GKG,  Facebook Graph Search delivers the powerful semantic search engine of Facebook, providing user-specific answers through the dynamic Facebook social knowledge base.

\subsubsection{Commonsense KGs}
\
\par

Commonsense knowledge graphs depict widely-accepted knowledge of common understandings. OpenCyc \cite{DBLP:conf/aaaiss/MatuszekCWD06} is one of the earliest of these attempts, which encodes knowledge concepts, rules, and common sense ideas in the form of CycL. Besides OpenCyc, ASER \cite{DBLP:journals/corr/abs-2104-02137} provides a weighted knowledge graph that describes commonsense by modeling entities of actions, states, events, and relationships among these objects, which acquire its nodes via dependency patterns selection and conceptualized by Probase. TransOMCS \cite{DBLP:conf/ijcai/ZhangKSR20} develops an auto-generated dataset covering 20 commonsense relations obtained from linguistic graphs.

\begin{table}[!t]
\centering
\caption{The infomation of off-the-shelf knowledge graph tools.}
\scalebox{0.75}{
\begin{tabular}{llll}
\toprule[1.5pt] 
Task & Tool & Year & URL\\
\midrule[1.5pt]  

\multirow{2}{*}{Data Pre-processing} & WebCollector&2016&https://github.com/CrawlScript/WebCollector\\ 
& Web Scraper (v0.4.0) & 2019 & https://webscraper.io/ \\ 

\hline
\multirow{13}{*}{Knowledge Acquisition} &
NLTK&2002&https://www.nltk.org/\\
& StanfordNLP&2002&https://stanfordnlp.github.io/stanfordnlp/\\
&KnowItAll&2005&https://github.com/knowitall\\
& TextRunner&2007&https://www.cs.washington.edu/research/textrunner/\\
& OpenCalais&2008&http://www.opencalais.com\\
& ReVerb&2011&https://github.com/knowitall/reverb\\
& OLLIE&2012&https://knowitall.github.io/ollie/\\

& spaCy&2017&https://spacy.io/\\
& TableMiner+&2017&https://github.com/ziqizhang/sti\\
& MantisTable&2019&http://mantistable.disco.unimib.it/\\
& OpenNRE&2019&https://github.com/thunlp/OpenNRE\\
& gBuilder&2021&http://gbuilder.gstore.cn/\\
\hline

\multirow{4}{*}{Knowledge Acquisition} & Falcon-AO&2008&http://ws.nju.edu.cn/falcon-ao/\\
& OpenKE&2018&https://github.com/thunlp/OpenKE\\
& OpenNE&2019&https://github.com/thunlp/OpenNE\\
& OpenEA&2020&https://github.com/nju-websoft/OpenEA\\

\bottomrule[1.5pt] 
\end{tabular}
}
\label{tabKGTools}
\vspace{-0.3cm}
\end{table}

\subsubsection{Domain-specific KGs}
\
\par

Researchers also assemble specific knowledge to serve multiple professional research fields. As for the biomedical field, Pubmed \cite{roberts2001pubmed} provides an open biomedical literature database released by the Pubmed center, while Durgbank \cite{DBLP:journals/nar/WishartKGSHSCW06} provides an insight into pharmacology with protein and drug information. Many KG collection efforts are also contributed to fighting against the COVID-19 pandemic, such as the COVID-19 Concept dataset \footnote{http://openkg.cn/dataset/covid-19-concept} and Aminer COVID-19 Open Data \footnote{https://www.aminer.cn/data-covid19/}. As for academic activities, the Academic Social Network (ASN) of AMiner \cite{DBLP:conf/icdm/TangZY07} discloses the scholars and their academic activities with the network containing paper and citation relationships. Similarly, Open Academic Graph (OAG) \cite{DBLP:conf/kdd/ZhangLTDYZGWSLW19} delivers integrated academic social networks with paper data. Creatively, Huapu \cite{xin2020huapu} builds a high-quality Chinese family book semantic network from digitized thousands of ancestral genealogy books, while providing relationship links between the same people in the different family trees established by fusing multiple Web sources.

\subsubsection{Federated KGs}
\
\par

Privacy protection plays a critical role for knowledge bases with massive user data from different providers, while knowledge communities collecting multi-source KG data possess abundant features, which can be unified to build integrated knowledge models. Federated learning is proposed to combine sub-KG features from different data providers with protection to prevent data exchange. Federation strategies have been applied to KG systems with sensitive data. GEDmatch\footnote{https://www.gedmatch.com/} is a genealogy KGs collecting user-provided information for enquiring DNA while distributing a federated knowledge model to probe data tracking. Federation strategies have been applied to many knowledge graph systems with sensitive data. These have helped to build integrated knowledge models while preventing data exchange. Researchers have also focused on federated knowledge graph platforms. OpenKG.cn \cite{DBLP:journals/dint/ChenHQWBLY21} is a crowd-sourcing community that provides a knowledge-sharing platform to develop knowledge applications with federated learning while supporting the decentralization of knowledge blockchains.

\subsection{Knowledge Graph Construction Tools}

In this section, we review some of the tools commonly used for knowledge graph construction. These tools mainly support construction sub-tasks, such as pre-processing, knowledge acquisition, and knowledge refinement. The details of these tools are displayed in Table \ref{tabKGTools}.

\subsubsection{Data Preprocessing}
\
\par
Data pre-processing tasks remove noise like advertisements. For instance, WebCollector \cite{gong2016online} is a representative data pre-processing toolkit that automatically filters non-content noise, such as advertisements and layout information, while retaining the page content via integrated algorithms. Many web crawlers also support data preprocessing tasks that extract informative structures or content. Beyond WebCollector, Web Scraper \footnote{https://webscraper.io/}  is a user-friendly manual extraction tool for collecting multiple web pages. It provides a user interface to preserve focused web structures and a cloud server for extracting massive amounts of content.

\subsubsection{Knowledge Acquisition}
\
\par

The early toolkits for knowledge acquisition directly extract fact triples through rules, patterns, and statistical features. These are also known as information extraction toolkits. Beyond KnowItAll \cite{DBLP:conf/www/EtzioniCDKPSSWY04}, other toolkits, such as TextRunner \cite{DBLP:conf/naacl/YatesBBCES07}, ReVerb \cite{ReVerb2011},  leverage semi-supervised designs for collecting relational information. These tools produce refined verbal triples via syntactic and lexical information. In addition, there is OLLIE \cite{DBLP:conf/emnlp/MausamSSBE12} which supports the discover of non-verbal triples.

Many NLP applications can directly perform knowledge acquisition sub-tasks, including named entity recognition, relation extraction, and coreference resolution tasks, or they can provide linguistic features for related applications. NLTK \cite{DBLP:journals/corr/cs-CL-0205028} and StanfordNLP \cite{DBLP:conf/acl/ManningSBFBM14} are powerful toolkits for knowledge acquisition based on statistical algorithms like conditional random field and MEM. These tools can also provide background features such as POS tags and NP chunks. Meanwhile, TableMiner+ \cite{DBLP:journals/semweb/Zhang17} and MantisTable \cite{DBLP:conf/esws/CremaschiRSP19} extract knowledge from semi-structured table forms.

Recently, developers have been drawn into deep learning-based toolkits. For example, spaCy \cite{spacy2} is a comprehensive practical NLP toolkit that integrates NeuralCoref for coreference resolution tasks and also provides a trainable deep learning module for specialized relation extraction (in spaCy v3). OpenNRE \cite{han-etal-2019-opennre} provides various extensible neural network models such as CNN and LSTM to perform supervised relation extraction. Deep learning toolkits provide high-performance techniques for users. For industrial applications, users can customize their knowledge acquisition solution by OpenCalais \footnote{http://www.opencalais.com}. Further, gBuilder \footnote{http://gbuilder.gstore.cn/} is a milestone end-to-end solution based on user-selected neural architectures that also supports an active learning interface for performance optimization. gBuilder can directly output a raw knowledge graph from inputted unstructured data.

\subsubsection{Knowledge Refinement}
\
\par

Knowledge refinement tools refine an existing raw knowledge graph by completing it or merging it with other knowledge graphs. Tools based on deep learning are popular solutions. For example, OpenKE \cite{DBLP:conf/emnlp/HanCLLLSL18} provides multiple knowledge representation models for knowledge graph completion, while OpenEA \cite{DBLP:journals/pvldb/SunZHWCAL20} leverages similar methods for merging structures in knowledge graphs. Beyond the integrated deep learning toolkits, OpenKE  and OpenEA, OpenNE\footnote{https://github.com/thunlp/OpenNE} integrates embedding models such as Node2VEC \cite{grover2016node2vec} and LINE \cite{tang2015line} to generate global representations from a complete knowledge graph. In terms of knowledge fusion (merging) tasks, Falcon-AO \cite{DBLP:journals/ws/HuQ08} uses multiple algorithms to measure semantic similarity so as to align concepts in different notations.

\section{Semi-structured Data  Pre-Processing}
\label{Section 3}
Real-world raw data sources include multi-structure contents with irrelevant parts impairing the effect of knowledge extraction. Data preprocessing is necessary for handling a messy data environment. Preprocessing sub-tasks mainly include Content Extraction and Structure Interpretation. 

\subsection{Content Extraction}
Many web pages contain non-content noises such as advertisements. Content extraction tasks aim to erase these irrelevant elements while reserving knowledge content. Users can manually select the main part of a web page, e.g.,  contents enveloped by ``<table>'', to achieve this goal by web crawlers such as JSoup, BeautifulSoup, and Web Scraper that retrieve and interpret elements in Document Object Model (DOM) structures, then users can select the main part of a web page. However, when the data volume is high, manual work will fail to handle them. Mainstream automatic content extraction methods mainly include wrapper-based methods and statistic-based methods.

Wrapper-based methods are the earliest attempts to detect main contents, leveraging matching rules to capture informative content. Off-the-shelf wrapper tools automatically generating rules from semi-structured pages 
include IEPAD \cite{DBLP:journals/dss/ChangHL03}, SoftMealy \cite{DBLP:journals/is/HsuD98}. Bootstrapping methods iteratively enhance extraction templates with seed examples, such as \cite{DBLP:conf/cikm/GolgherSLR01} and  \cite{DBLP:conf/pkdd/CarlsonS08}. Some toolkits providing user interfaces to optimize extraction templates, such as NoDoSE \cite{DBLP:conf/sigmod/Adelberg98} and  DEByE \cite{DBLP:journals/dke/LaenderRS02}. Template-based wrappers are easy to understand and achieve feasible results where the page structures are well-formed, but fail to grasp the inner contents covered by intricate novel elements or structures.

Users can also utilize methods based on statistical features of web pages to obtain informative content. Finn et.al \cite{DBLP:conf/delos/FinnKS01} propose an empirical assumption that an informative sub-sequence in a web page contains sufficient enough words with minimal tags. Many models consider statistical features of web contents for extracting informative content, such as  CETR \cite{DBLP:conf/www/WeningerHH10} (the ratio of text length to tag number),  CETD \cite{DBLP:conf/sigir/SunSL11} (text density in each sub-tree structure of a DOM tree) and  CEPR \cite{DBLP:conf/cikm/WuLHW13} (path ratio of Web links). Users can utilize WebCollector \cite{gong2016online} that integrates the statistic-based models for content extraction. Another heuristic research direction is visual-features-based methods. For example, VIPS \cite{DBLP:conf/apweb/CaiYWM03} utilizes the visual appearances (such as fonts and color  types) of a page to build a content structure tree for content extraction.

When content extraction has been performed on a semi-structured page, users will acquire a renewed noise-free semi-structured or unstructured document.

\subsection{Structure Interpretation}
Many table forms in the web pages function as navigators or style-formatted containers for contents (handled by content extractors), comprising no relational structures. Models shall filter these decorative non-relational web table structures before obtaining relational information.

Relational table interpretation is a binary classification task that determines whether a table is informative. Methods analyze semantic features of table structures for classification. Wang and Hu \cite{DBLP:conf/www/WangH02} design a table classifier integrated with support vector machines (SVM) and decision trees based on the layout and content type features. Similarly, WebTables \cite{DBLP:conf/webdb/CafarellaHZWW08} develops a rule-based classifier based on the table size (number of rows and columns) and tags. Eberius et al. \cite{DBLP:conf/bdc/EberiusBHTAL15} develop a classification system DWTC via the feature of the table matrixes. Many web tables also contain data noises. OCTOPUS \cite{DBLP:journals/pvldb/CafarellaHK09} further incorporates data cleansing with table classification tasks to filter informative tables.

Developing a table interpretation model includes two steps: first select features in the table forms, then integrate learning models to analyze relational semantics in the data. We recommend readers to refer to \cite{DBLP:journals/tist/ZhangB20} for more table syntax features and high-performance model ensembles. 

\begin{figure*}[!t]
 \centering
 
 \includegraphics{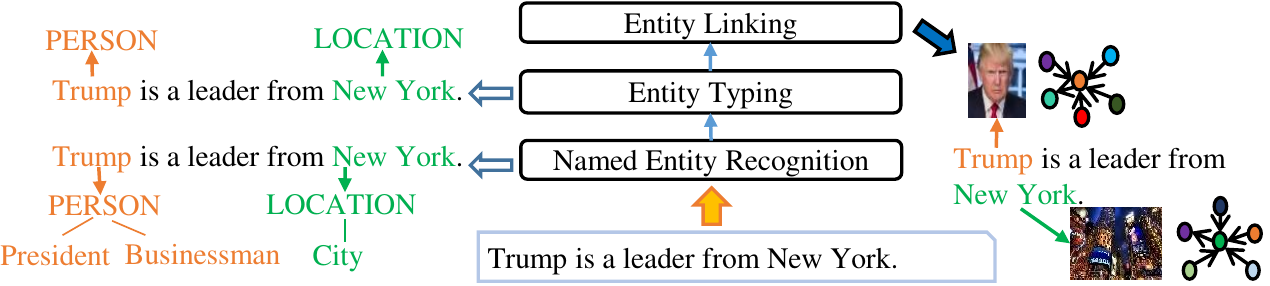}
 \caption{The entity discovery process.}
 \label{fig:figure41}
 \vspace{-0.34cm}
\end{figure*} 

\section{Knowledge Acquisition} \label{Section 4}
Knowledge acquisition is the general process of collecting elements from multi-structured data to build a knowledge graph. It includes entity recognition, coreference resolution, and relation extraction. Entity recognition tasks discover entity mentions within data. Co-reference resolution tasks then locate referred mention pairs, followed by relation extraction tasks, which link entities with their semantic relationships.

\subsection{Entity Discovery}

Entity discovery acquires a subset of concepts from semi-structured or structured data that can constitute the nodes of a knowledge graph. The general procedure of entity discovery includes named entity recognition, entity typing, and entity linking tasks. Named entity recognition tasks discover strings that refer to semantic entities and then classify them to the general types (e.g., person, location, country, company). Entity typing tasks categorize the found entities into specific types (e.g., actor, artist, brand). Entity linking associates a discovered entity with a possible node in the knowledge graph. If there are no available nodes for linking, a corresponding entity node will be created to represent the newly found entity. Fig. \ref{fig:figure41} depicts an overview of the general process.

\subsubsection{Named Entity Recognition from semi-structured data vs unstructured data}
\
\par


Named entity recognition tasks tag named entities in semi-structured or unstructured data with their positions and classifications. Semi-structured data are enveloped by semantic hints related to property-attribute structures, while unstructured data only contains texts.

Rule-based approaches \cite{kushmerick2000wrapper} are the general solutions for NER. As for semi-structured web data, Wrapper inductions generate rule wrappers to interpret semi-structures such as DOM tree nodes and tags for harvesting entities from pages. Some rule-based solutions are unsupervised, which requires no human annotations, such as Omini \cite{DBLP:conf/icdcs/ButtlerLP01}. As for entities in table forms, many approaches are proposed based on property-attribute layouts of Wikipedia, such as rule-based tools \cite{DBLP:conf/semweb/AuerBKLCI07}\cite{DBLP:conf/www/SuchanekKW07} for DBpedia, and YAGO.   For unstructured data, classic NER systems \cite{DBLP:conf/tipster/Sundheim96} also rely on manually-constructed rule sets for pattern matching. Semi-supervised approaches are developed to improve rule-based NER by iteratively generating refined new patterns via pattern seeds and scoring, such as Bootstrapping-based NER \cite{DBLP:journals/corr/ThenmalarBG15}.

\begin{figure*}[!t]
 \centering
 \includegraphics{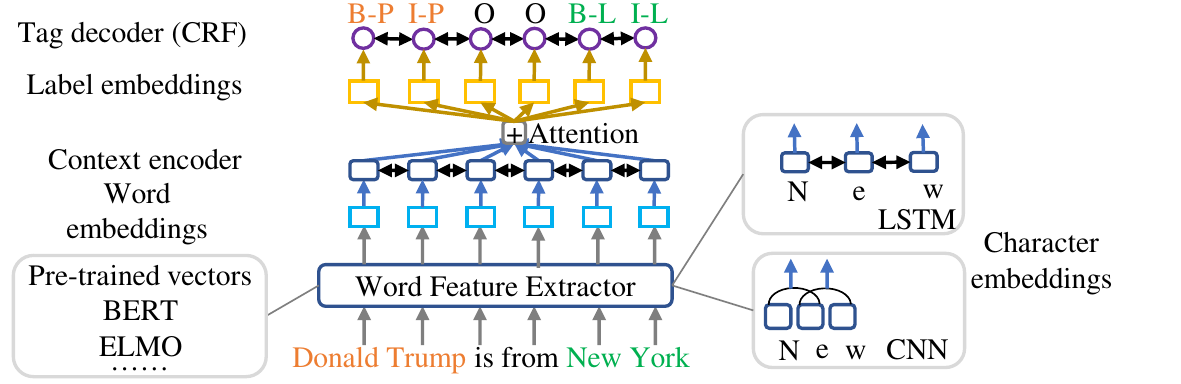}
\scriptsize
 \caption{Illustration of the standard architecture for deep-learning-based named entity recognition. When a sentence is input, such NER model will output tagged entity words with positional information and rough entity classifications.}
 \label{fig:figure4111}
 \vspace{-0.76cm}
\end{figure*} 

Statistic-based approaches treat named entity recognition as a sequential classification tagging task that tags entities according to the BIES scheme (beginning, intermedia, ending, single) and their types. For unstructured named entity recognition, the key hypothesis is that a tag for each word only depends on the previous words. Hence, applications built on hidden Markov models \cite{DBLP:conf/acl/ZhouS02} and conditional random fields (CRF) models \cite{DBLP:conf/acl/FinkelGM05}, which capture neighborhood-dependencies, are popular NER designs. With semi-structured table data, researchers often use CRF variants to tackle the two-dimensional features of the attributes that relate to the entities, such as 2D-CRF \cite{DBLP:conf/icml/ZhuNWZM05}. Thus, they extract multiple attributes of each entity in a 2D structure. Dynamic conditional random field (DCRF) \cite{DBLP:conf/icml/SuttonRM04} infers potential attribute-entity interaction via the dynamic Bayesian network,  and hierarchical CRF  \cite{DBLP:conf/kdd/ZhuNWZM06} models semi-structured data into a hierarchical tree for joint extraction. Further, Finn and Kushmerick \cite{DBLP:conf/ecml/FinnK04} developed a model to locate the boundaries of entities in texts which is based on SVM.

Deep learning is also becoming a popular trend in named entity recognition, especially for text named entity recognition. These deep learning approaches typically treat named entity recognition as a seq2seq model (words sequences to label sequences). These models aggregate contextual embeddings according to the input, and context encoders then output word type tags through tag decoders such as a CRF structure or a softmax structure \cite{DBLP:journals/tkde/LiSHL22}.

CNN structures mainly focus on local features for capturing entities. Colloabert et al. \cite{DBLP:journals/jmlr/CollobertWBKKK11} was the first to employ a CNN with a CRF output layer as a unified solution for entity detection. IDCNN et al. \cite{DBLP:conf/emnlp/StrubellVBM17} improved upon the CNN with dilated convolutions that enlarge the perception field by omitting some of the input to enhance generalization.

RNN structures can better digest global contextual features in long sentences, such as the uni-directional RNN for biomedical entity recognition presented in \cite{DBLP:conf/bibm/LiJJSH15}. However, RNNs may suffer from context bias with later upcoming words \cite{DBLP:journals/corr/HuangXY15}. Hence, many models consider bi-directional RNNs, such as the Bi-LSTM-CRF-based in \cite{DBLP:journals/corr/HuangXY15} and the GRU-based NER model in \cite{DBLP:journals/corr/NguyenSDF16}. Combinations of character and word encoders are also widely-applied structures, such as a structure comprising a CNN for character embedding and an LSTM for word embedding \cite{DBLP:conf/acl/MaH16}. The standard architecture of deep learning-based NER models is pictured in Fig. \ref{fig:figure4111}.

Another direction that projects salient interactions for global contexts is the attention mechanism. Luo et al. \cite{DBLP:journals/bioinformatics/LuoYYZWLW18} introduce word-level soft attention to enhance named entity recognition. Gregoric et al. \cite{DBLP:conf/ictai/GregoricBMCM17} employ word-word self-attention for named entity recognition.

Graph convolution networks are often used to handle context in linguistic graph structures for named entity recognition. For example, Cetoli et al. \cite{DBLP:conf/tlt/CetoliBOS18} proposed a GCN framework that encodes the LSTM-proceed features via GCN structures with a syntactic dependency tree. Pre-trained language models that provide representations as background knowledge for training with named entity recognition tasks have also achieved breakthroughs in named entity recognition. Models include Elmo \cite{DBLP:journals/corr/abs-1904-10503}, Ltp \cite{DBLP:journals/corr/abs-2001-02524}, and LUKE \cite{DBLP:conf/emnlp/YamadaASTM20}.

\begin{figure*}[!t]
 \centering
 \includegraphics{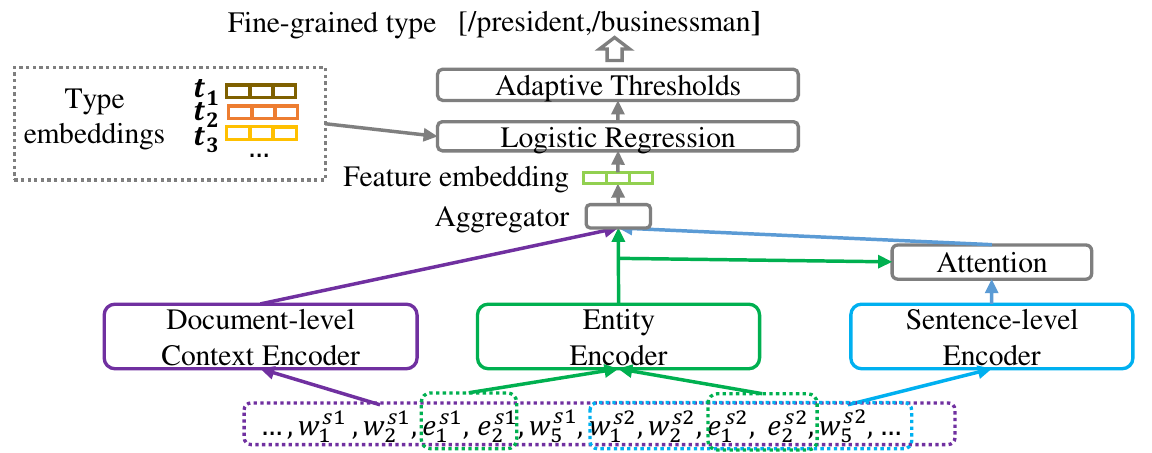}
 \caption{ Illustration of deep learning-based entity typing via multi-scale feature extraction (based on \cite{DBLP:conf/starsem/ZhangDD18}).}
 \label{fig:figure4121}
 \vspace{-0.4cm}
\end{figure*} 
 
\subsubsection{Entity Typing}
\label{et}
\
\par

Entity typing (ET) tasks provide fine-grained and ultra-grained type information for entities such as scientists, clubs, and hotels. Information loss occurs if ET tasks are not performed, e.g., Donald Trump is a politician and a businessman. Semi-structured tables provide hints for fine-grained types in the captions. For example, ``soccer players in England'' suggests soccer players as the entity. However, tagging proper fine-grained entity types in different contexts for unstructured data can be intricate.

Deep learning approaches tackle two main challenges for entity typing: 1) infrequent fine-grain types; and 2) overly-specific typing. Some specific types can be imbalanced or infrequent. For this reason, Shimaoka et al. \cite{DBLP:conf/eacl/InuiRSS17} proposed an LSTM-based attentive neural network for infrequent entity typing that relies on hierarchical label encoding integrated with mention and context representations to exploit fine-grained contextual features. Overly-specific type annotations derive correct types but do not fit the current data context. Xu et al. \cite{DBLP:conf/naacl/XuB18} applied an out-of-context loss function to the entities with multiple labels for filtering overly-specific data noise which assumed that the type label which scored the highest probability during training was correctly tagged. To further explore context scenarios, Zhang et al. \cite{DBLP:conf/starsem/ZhangDD18} introduced document-level representations to provide a global context for discovering entities. Sentence-level contextual representations are then used to align the same entity representations appearing in different sentences. An adaptive probability threshold then generates labels of the entity types in different contexts. Fig. \ref{fig:figure4121} presents a typical deep learning-based ET model.

\begin{figure*}[!t]
  \centering
  
  \includegraphics[width=\linewidth]{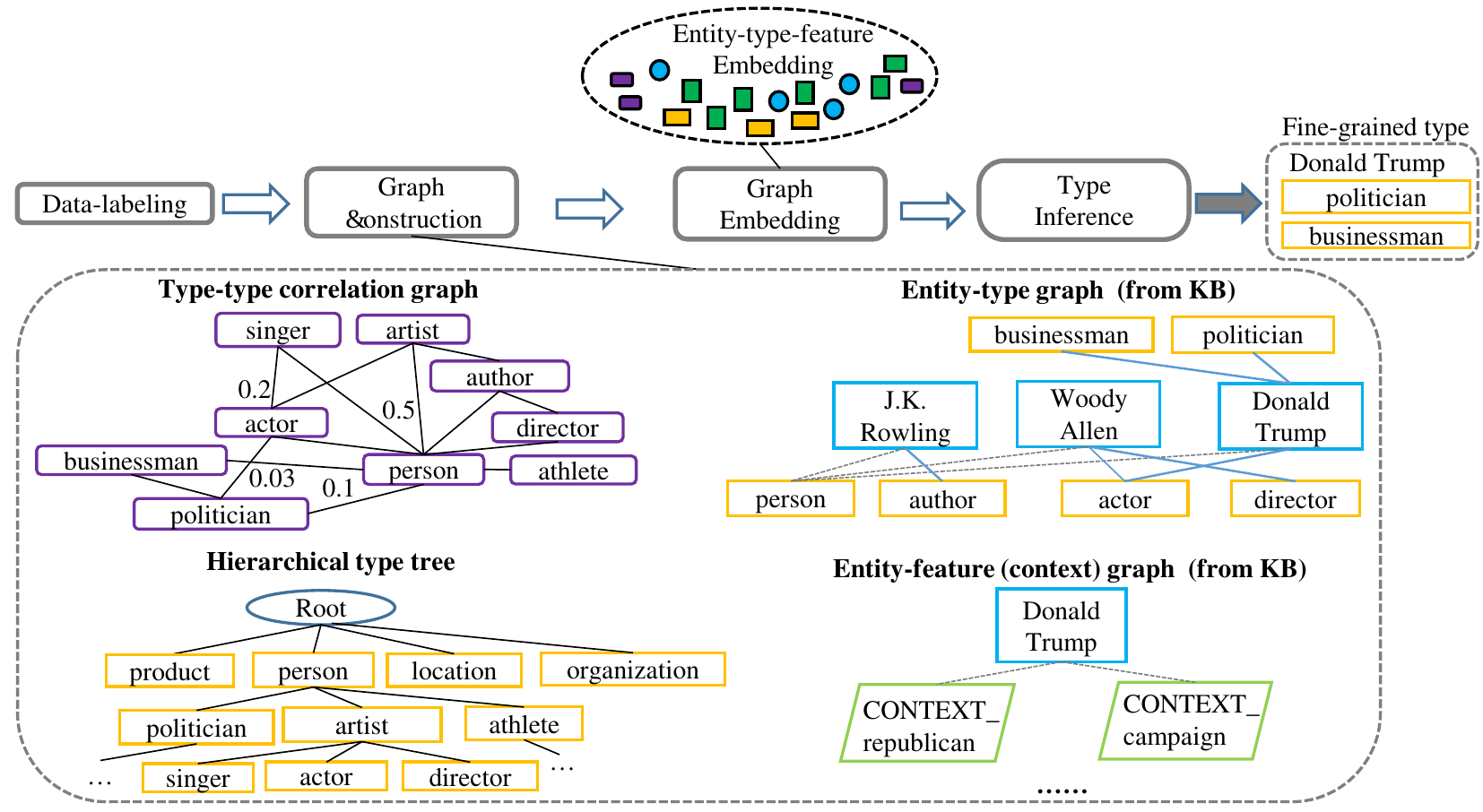}
  \caption{Illustration of embedding-based ET via heterogeneous graph structures (PTE \cite{DBLP:conf/kdd/RenHQVJH16}).}
  \label{fig:figure4122}
  \vspace{-0.38cm}
\end{figure*}

 \label{etlcontext}
 Novel embedding-based models avail of combing global graph structure features and background knowledge for predicting potential types of entities via representations. Researchers reported that the classical TransE model acts poorly while directly applied to ET tasks. Moon et al. \cite{DBLP:conf/cikm/Moon0S17} propose the TransE-ET model adjusting the TransE model by optimizing the euclidean distance between entities and their types representations, limited by insufficient entities types and triples features. New solutions aim at constructing various graphs to share diversified features of entity-related objects for learning embeddings with entity-type features. PTE \cite{DBLP:conf/kdd/RenHQVJH16} reduces data noise via a partial-label embedding, which constructs a bipartisan graph between entities and all their types while connecting entities nodes to their related extracted text features. Finally, PTE utilizes the background KG by building a type hierarchy tree with the derived correlation weights.  JOIE \cite{DBLP:conf/kdd/HaoCYSW19}  embeds entity nodes in the ontology-view graph and instance graphs, gathering entity types by top-k ranking between entity and type candidates. Likewise, ConnectE \cite{DBLP:conf/acl/ZhaoZXLW20} maps entities onto their types and learning knowledge triples embeddings. Practical models improving embeddings on heterogeneous graphs for ET tasks (in Xlore project \cite{DBLP:conf/semweb/WangLWLLZSLZT13}) also include \cite{DBLP:conf/cncl/JinHL18}, \cite{DBLP:conf/coling/JinHLD18}, \cite{DBLP:conf/emnlp/0002HLLLCD18}. We present graph structures for embedding model-based ET in Fig. \ref{fig:figure4122}.

\subsubsection{Entity Linking from semi-structured data vs unstructured data}
\
\par

Entity linking (EL) tasks, also called entity disambiguation, link entity mentions to their corresponding objects in a knowledge graph. A textual mention can have different references, e.g., the text ``Tesla'' may refer to the car, the corporation, or the scientist. Entity linking connects mentions in different data backgrounds with the contextual information of their respective nodes. With semi-structured data, entity linking identifies entities using semantic hints from column heads, type labels, cell texts of tables, and hyperlinks. With unstructured text, entity linking models focus on the contextual representations of entity mentions.

Statistical approaches, especially those approaches based on probabilistic graphs and SVM models, are the general solutions for semi-structured and unstructured data. Models based on probabilistic graphs construct a probabilistic graph of mentions in tables, then link entities by calculating the semantic factors of nodes. Limaye et al. \cite{DBLP:journals/pvldb/LimayeSC10}, for instance, constructed a factor graph for collective entity linking based on the TF-IDF algorithm that calculates the term frequency of entity labels with cell-text pairs and type labels with column-head pairs. Some models incorporate external knowledge bases to improve entity linking tasks. For example, TabEL \cite{DBLP:conf/semweb/BhagavatulaND15} improves its factor graph by leveraging the hyperlinks in Wikipedia to estimate semantic relatedness features before collective classification for disambiguation. Wu et al. \cite{DBLP:conf/jist/WuYPXWQ16} propose an approach for enhancing entity linking with the ``same-as'' edges in multiple knowledge bases. Efthymiou et al. \cite{DBLP:conf/semweb/EfthymiouHRC17} systematically exploit semantic features for entity linking. Their approach integrates vector representations of an entity’s context, minimal entity context, and schematic structures shared between knowledge bases and tables. SVM models treat entity linking as a classification task. Here, Mulwad et al. \cite{DBLP:conf/semweb/MulwadFSJ10a} develop a model based on SVMRanker that determines which potential nodes can link to a target entity. Similarly, Guo et al. \cite{DBLP:conf/ijcnlp/GuoCLL11} propose a probabilistic model for unstructured data, that leverages the prior probability of an entity, context, and name when performing linking tasks with unstructured data. Han et al. \cite{DBLP:conf/sigir/HanSZ11} employed a reference graph of entities, assuming that entities co-occurring in the same documents should be semantically related.

\begin{figure*}[!t]
 \centering
 \includegraphics{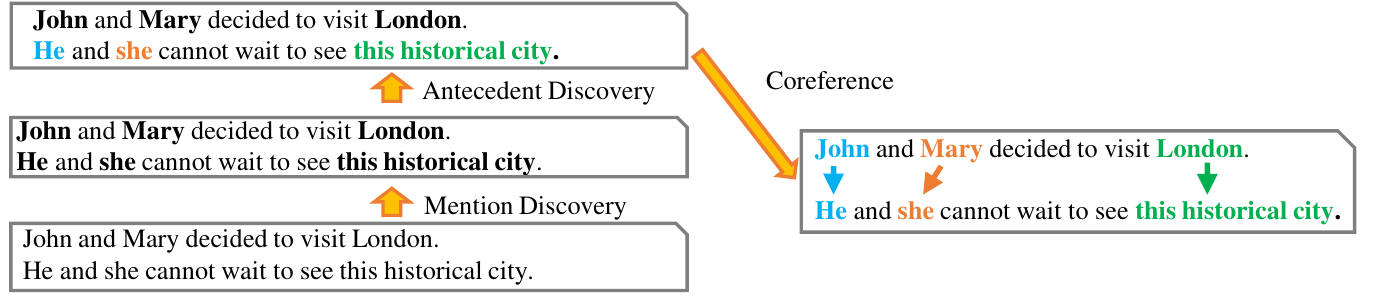}
\scriptsize
 \caption{The coreference resolution process. First, mentions are detected. Then the antecedents of the mentions are selected and matches to co-referred pairs. Noticeably, coreference resolution tasks can be performed on documents with multiple sentences, while handling identical mentions in a compound sentence.}
 \label{fig:figure42}
 \vspace{-0.38cm}
\end{figure*} 

Embedding-based models are also critical solutions for entity linking  via entity embeddings.  LIEGE \cite{DBLP:conf/kdd/ShenWLW12}  derives distribution context representations to links entities for web pages.  Early researchers \cite{DBLP:conf/acl/BaggaB98} leverage Bag-of-word (BoW) for contextual embeddings of entity mentions, then performed clustering to gather linked entity pairs. Later, Lasek et al. \cite{DBLP:conf/iiwas/LasekV12} extend the BoW model with linguistic embeddings for EL tasks. Researchers also focus on Deep representations for high-performance linking. DSRM \cite{DBLP:journals/corr/HuangHJ15} employs a deep neural network to exploit semantic relatedness, combining entity descriptions and relationships with types features to obtain deep entity features for linking.  EDKate \cite{DBLP:conf/conll/FangZWCL16} jointly learns low-dimensional embedding of entities and words in the knowledge base and textual data,  capturing intrinsic entity-mention features beyond the BoW model. Furthermore, Ganea and Hofmann \cite{DBLP:conf/emnlp/GaneaH17} introduce an attention mechanism for joint embedding and passed semantic interaction for disambiguation. Le and Titov \cite{DBLP:conf/acl/TitovL18a} model the latent relations between mentions in the context for embedding, utilizing mention-wise and relation-wise normalization to score pair-wise coherence score function.

\subsubsection{Other Advances}
\
\par

Few/Zero-shot entity typing is an intricate challenging issue. Ma et al. \cite{DBLP:conf/coling/MaCG16} develops Proto-HLE that models the prototype of entity label embeddings for zero-shot fine-grain ET tasks, combining prototypical features with hierarchical type labels for inferring essential features of a new type. Zhang et al. \cite{DBLP:conf/coling/ZhangXLY20} further propose MZET that exploits contextual features and word embeddings with a Memory Network to provide semantic side information for few-shot entity typing.
 
 Joint extraction models incorporating NER with EL tasks reduce error propagation of the pipeline-based entity recognition tasks. NEREL \cite{DBLP:conf/cikm/SilY13} couples NER and EL tasks by ranking extracted mention-entity pairs to exploit the interaction features between entity mentions and their links. Graphic models are also effective designs to combine NEN (Named Entity Normalization) labels that convert entity mentions into unambiguous forms, e.g., Washington (Person) and Washington (State).  Liu et al. \cite{DBLP:conf/acl/LiuZZFW12} incorporated EL with NEN tasks utilizing a factor graph model, forming CRF chains for word entity types and their target nodes. Likewise, MINTREE \cite{DBLP:journals/tkde/PhanSTHL19} introduces a tree-based pair-linking model for collective tasks.

\clearpage

Researchers explore more strategies for flexible NER tasks. Transfer Learning shares knowledge between different domains or models. Pan et al. \cite{DBLP:journals/tois/PanTS13} propose Transfer Joint Embedding (TJE) to jointly embed output labels and input samples from different domains for blending intrinsic entity features. Lin et al. apply \cite{DBLP:conf/emnlp/LinL18} a neural network with adaptation layers to transfer parameter features from a model pre-trained on a different domain. Reinforcement Learning (RL) puts NER models to interact with the environment domain through a behavior agency with a reward policy, such as the Markov decision process (MDP-based) model \cite{DBLP:conf/emnlp/NarasimhanYB16} and Q-network enhanced model \cite{DBLP:journals/nature/MnihKSRVBGRFOPB15}.  Noticeably, researchers \cite{DBLP:conf/coling/YangCLHZ18} also leveraged the RL model for noise reduction in distant-supervised NER data. Adversarial Learning generates counterexamples or perturbations to enforce the robustness of NER models, such as DATNet \cite{DBLP:conf/acl/ZhouZJZFGK19} imposing perturbations on word representations and counterexamples generators (\cite{DBLP:conf/emnlp/Cao000L18}, \cite{DBLP:conf/ijcai/0034YS19}). Moreover, Active Learning, which queries users to annotate selected samples, has also been applied for NER. Shen et al. \cite{DBLP:conf/iclr/ShenYLKA18} incrementally chose the most samples for NER labeling during the training procedures to mitigate the reliance on tagged samples.

\begin{figure*}[!t]
 \centering
 \includegraphics[width=\linewidth]{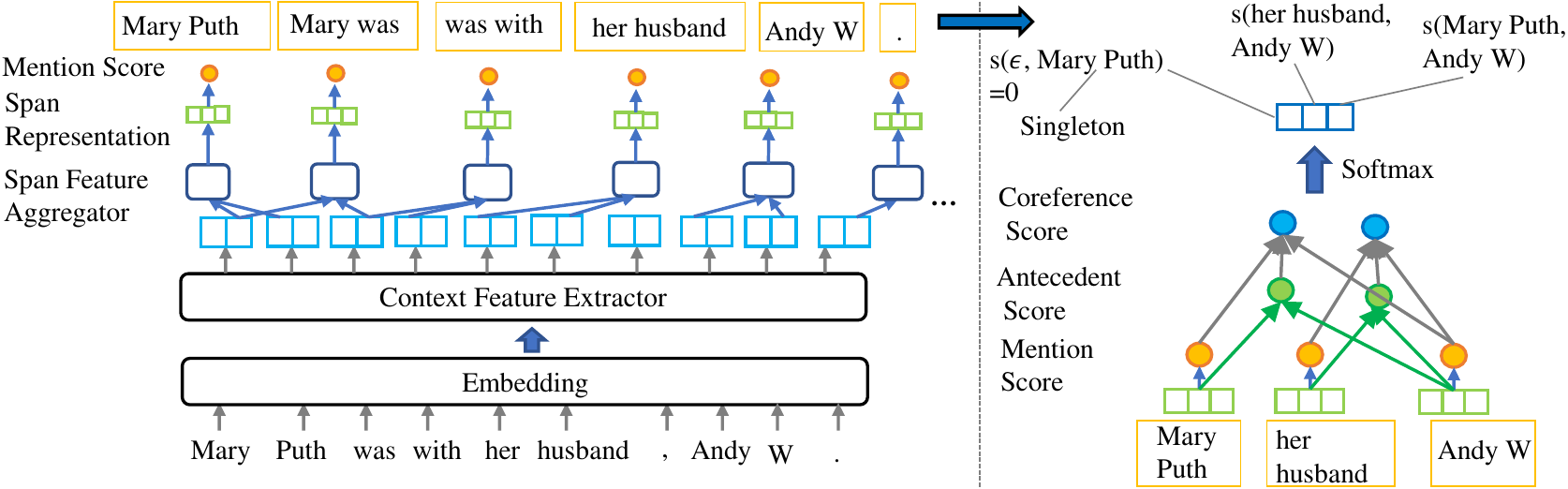}
 \scriptsize
 \caption{Architecture of the standard deep end-to-end model (based on \cite{DBLP:conf/emnlp/LeeHLZ17}). A deep learning model performs a two-stage procedure to tackle coreference resolution tasks: 1) Mention detection, which discovers entity mentions as spans from text; 2) Coreference detection, which score the antecedents in the span to match coreference mention pairs as outputs. Spans include combinations of all word sequences. This figure displays simplified results.}
 \vspace{-0.38cm}
 \label{fig:figure422}
\end{figure*} 

\subsection{Coreference Resolution} 
\label{CO}
Coreference expressions often appear in unstructured text. As such, coreference resolution (CO) tasks detect mentions that refer to the same entities (including aliases and pronouns). A mention will be a singleton if no other mentions refer to it. Given some unstructured sentences, such tasks will output co-referred word span pairs. Fig. 
 \ref{fig:figure42} presents this process.

\subsubsection{Statistic-based Models} 
\
\par

Early attempts to capture co-referred linguistic objects focused on the statistical features of entities, mentions, and antecedents.

 Cluster-based solutions handle the CO task as a pairwise binary classification task (co-referred or not). Early cluster models aim at mention-pair features. Soon et al. \cite{DBLP:journals/coling/SoonNL01} propose a single-link clustering strategy to detect anaphoric pairs. Recasens et al. \cite{DBLP:conf/naacl/RecasensMP13} further develop a mention-pair-based cluster to emanate a coreference chain or a singleton leaf. Later, researchers concentrate on entity-based features to exploit complex anaphoric features. Rahman and Ng \cite{DBLP:conf/emnlp/RahmanN09} propose a mention-ranking clustering model to dive into entity characteristics. Stoyanov and Eisner \cite{DBLP:conf/coling/StoyanovE12} develop agglomerative clustering to merge the best clusters with entity features.

\clearpage

Tree-based models and graph-based models are also popular designs for converting coreference resolution into a partition task. These models construct a hypergraph from the given document, in which each edge can link more than two nodes for modeling coreferences among multiple mentions. Cai and Strube \cite{DBLP:conf/coling/CaiS10} learn statistical features to weight edges and obtain coreference partitions via clustering algorithms. Sapena et al. \cite{DBLP:journals/coling/SapenaPT13} further employ relaxed labeling to interpret coreferences. Researchers have also simplified graphs for coreference resolution to adapt to tree-based methods. For example, Bean and Riloff \cite{DBLP:conf/naacl/BeanR04} introduced a decision tree model to distinguish anaphoric mentions combined with context features. Fernandes et al. \cite{DBLP:conf/conll/FernandesSM12} leverage a voted perceptron algorithm to detect mention-pair coreference trees.

\begin{figure*}[!t]
 \centering
 \includegraphics[width=\linewidth]{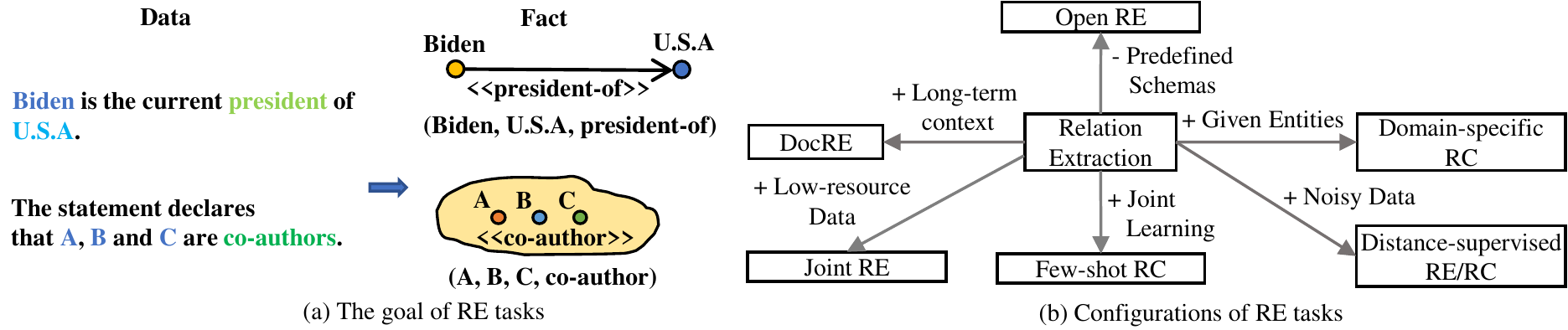}
 \caption{ Relation extraction. The goal of relation extraction tasks is to extract factual triples from data and find edges to link nodes together. For $n$-ary relations, the link is a super-edge that covers multiple nodes. }
 \label{fig:figure43}
\vspace{-0.38cm}
\end{figure*} 

 
\subsubsection{Deep Learning-based Models} 
\
\par

Deep learning models automatically convert a document input into word representations to collect features for detecting coreference mention pairs.

Many early models are based on CNNs. Xi et al. \cite{DBLP:conf/iscide/XiZHF15} resolve coreferences by incorporating distant features with hierarchical mention-pair features and also score mention pairs via a softmax layer. Wu et al. \cite{DBLP:conf/semco/WuM17} develop a CO model to effectively handle coreference and singleton expressions with abundant multi-scale contexts, incorporating context feature combinations of antecedents, mentions, and mention pairs features via convolution and concatenation.

RNN and its variants better extract global features between word mention pairs. Wiseman et al. \cite{DBLP:conf/naacl/WisemanRS16} propose an RNN-based CO model.  Lee et al. \cite{DBLP:conf/emnlp/LeeHLZ17} develop an end-to-end LSTM-based model, detecting internal dependencies within mentions spans to comprehend global contexts that surround the spans. Gu et al. \cite{DBLP:conf/cncl/GuLI18} apply a cluster modification algorithm to LSTM to rule out dissimilar pairs. Fig. \ref{fig:figure422} depicts a standard deep end-to-end CO model.

Attention mechanisms model semantic interactions for CO tasks. A good example of this is the Bi-LSTM structure enhanced with the word-level attention presented in \cite{DBLP:conf/emnlp/LeeHLZ17}. However, many different coreference resolution-specific attention mechanisms have been developed to exploit coreference features. These include: the biaffine attention model for CO tasks \cite{DBLP:conf/acl/ZhangSYXR18} that captures word span interactions for detecting linked expressions; and the mutual attention model \cite{DBLP:conf/wsdm/Ma0LHPSL20} that incorporates syntactic features with interactive features between dependency structures and antecedents for word spans. Further, Clark and Manning et al. \cite{DBLP:conf/emnlp/ClarkM16} employ RL-based  strategy to enhance the robustness of their neural  CO model, which uses a heuristic policy network to filter out wrong coreference matching actions.

Researchers also focus on embedding-based distribution models over multiple semantic structures to handle coreference resolution. Durrett and Klein \cite{DBLP:conf/emnlp/DurrettK13} utilize antecedents representations to enable coreference inference through distribution features. Martschat and Strube \cite{DBLP:journals/tacl/Martschat015} explore distribution semantics over mention-pairs and tree models to enhance coreference representations, directly picking robust features to optimize the CO task. Chakrabarti et al. \cite{DBLP:conf/kdd/ChakrabartiCCX12} further employ the MapReduce framework to cover anaphoric entity names through query context similarity.
\clearpage

\begin{figure*}[!t]
 \centering
 \includegraphics[width=\linewidth]{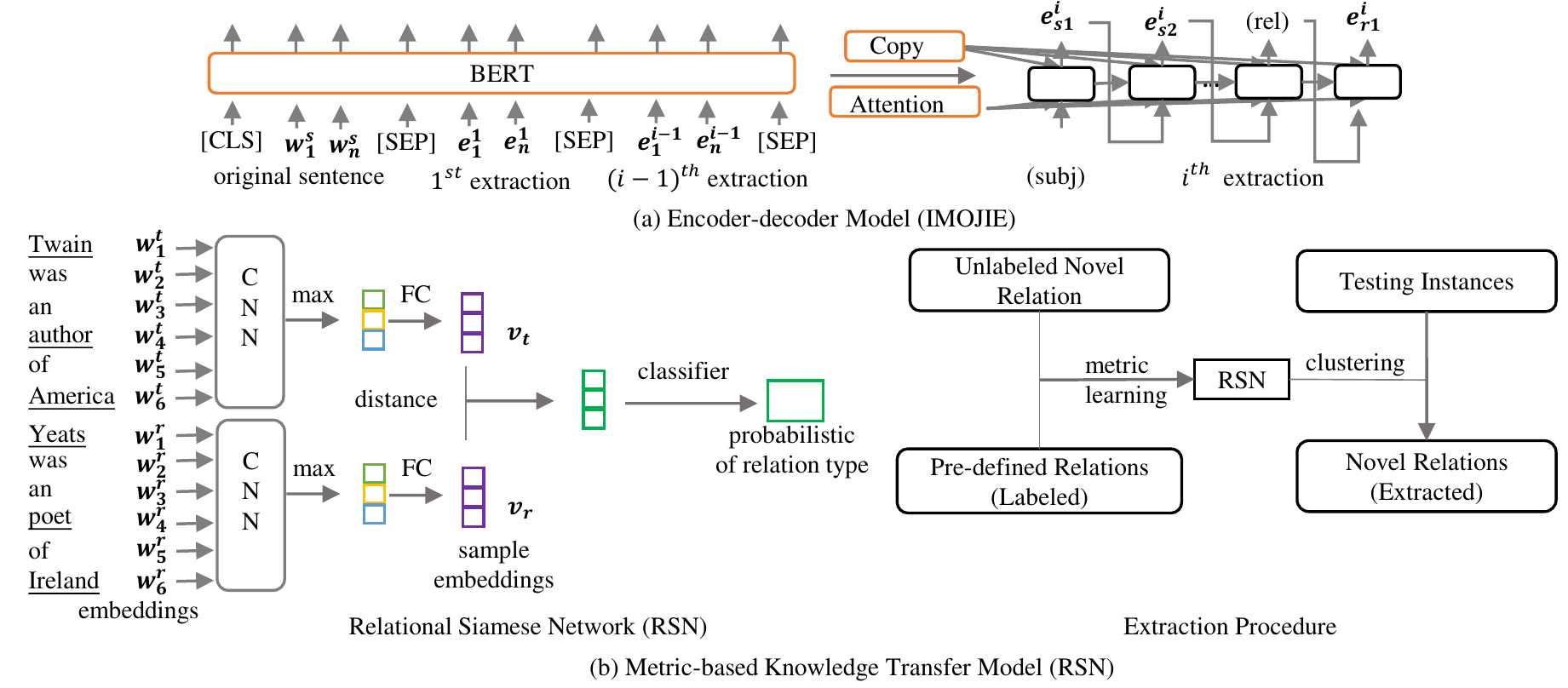}
 \caption{ Two paradigms of deep learning-based open relation extraction. In this figure, (a) shows the IMOJIE model \cite{DBLP:conf/acl/KolluruARMC20}, which extracts facts via a encoder-decoder design. (b) portrays a model \cite{DBLP:conf/emnlp/WuYHXLLLS19} that uses the RSN to compare relational patterns, then leverages clustering to collect the relations. }
 \label{fig:figure431}
 \vspace{-0.4cm}
\end{figure*}

\subsection{Relation Extraction}

Relation extraction tasks extract relational facts from unstructured or semi-structured data to indicate interactions and properties among entities. Relation extraction, as a downstream task, is often called relation classification. Binary relation extraction extracts relation triples between entity pairs, while n-ary relation extraction obtains relation triples over multiple entities, such as co-authors. Relation extraction endows a knowledge graph with semantic 
links. Fig. \ref{fig:figure43} presents an overview of the relation extraction tasks.

\subsubsection{Open Relation Extraction from semi-structured data vs unstructured data} 
\
\par

Open relation extraction tasks discover facts from unstructured data without pre-defined relation types. These techniques detect nominal words (as the subject or object) and verbal phrases (as the predicate) from free text to form knowledge triples like (subject, predicate, object).

Statistical approaches are also trending solutions for open relation extraction. In terms of relation detection, models based on probabilistic-graph are popular designs for allowing contextual information to flow through semi-structured structures or unstructured free text. Mulwad et al. \cite{DBLP:conf/semweb/MulwadFJ13} put forward a probabilistic graph on semi-structured tables and semantic message passing for tagging relationships. Chen and Cafarella \cite{DBLP:conf/vldb/ChenC13} leveraged a module based on the CRF structure with a frame finder to tag cells with their location labels (such as left, middle, and right). From this, a hierarchical tree is built where relation triples can be recovered through parent-child structures. Researchers have also applied probabilistic models to relation classification with text. As an example, StatSnowball \cite{DBLP:conf/www/ZhuNLZW09} employs Markov logic networks to identity relationships.

Methods focusing rules are the earliest attempts for RE tasks on different data structure kinds, gathering strings that fit in hand-craft templates, e.g., ``\$PEOPLE is born in \$LOCATION.'' refers to (\$PEOPLE, born-in, \$LOCATION). However, these unsupervised strategies rely on complex linguist knowledge to label data. Later, researchers concentrate on automatical pattern discovery for triples mining. Semi-supervision design is an enlightening strategy to reduce hand-craft features and data labeling that uncovers more reliable patterns based on a small group of annotated samples, such as DIPRE \cite{DBLP:conf/webdb/Brin98} iteratively extracting patterns with seeds, bootstrapping-based KnowItAll \cite{DBLP:conf/www/EtzioniCDKPSSWY04} and Snowball \cite{DBLP:conf/dl/AgichteinG00} equipping DIPRE with confidence evaluation. Some rule-based models consider more lexical objects for mining. OLLIE \cite{DBLP:conf/emnlp/MausamSSBE12} incorporates lexical structure patterns with relational dependency paths in texts. MetaPAD \cite{DBLP:conf/kdd/JiangSCRKH017} combines lexical segmentation and synonymous clustering to meta patterns that are sufficiently informative, frequent, and accurate for relational triples. Specifically for semi-structured tables, researchers design table structure-based rules to acquire relationships 
arranged in rows, columns, and table headers, such as \cite{DBLP:conf/kbse/AhmadAGK03}. Furthermore, Some semi-structured extraction systems utilizing distant supervision tolerate potential errors, which directly query external databases like DBpedia and Wikipedia to acquire relationships for the found entities in tabular data, such as the previous methods \cite{DBLP:conf/semweb/MulwadFSJ10a}, \cite{DBLP:conf/semweb/MulwadFJ13}, and  \cite{DBLP:conf/www/SekhavatPBM14}. Similarly, Muñoz et al. \cite{DBLP:conf/wsdm/MunozHM14} look up the Wikipedia tables for labeling relationships in tabular forms. Krause et al. \cite{DBLP:conf/semweb/KrauseLUX12} also expand rule sets for relation extraction via distant supervision.

Deep learning models have also been developed to handle open relation extraction. A common framework is an encoder-decoder model designed to acquire factual patterns. CopyAttention \cite{DBLP:conf/acl/CuiWZ18} includes a mechanism to copy words from input to output sequences via a neural bootstrapping strategy. IMOJIE \cite{DBLP:conf/acl/KolluruARMC20} improves CopyAttention with BERT-LSTM structures while incorporating an unsupervised aggregation scheme to perform iterative extraction. Another direction in open relation extraction is to transfer supervised knowledge to a model so as to adapt known relations to obtain unsupervised relations. In this vein,  Wu et al. \cite{DBLP:conf/emnlp/WuYHXLLLS19} developed a metric learning-based solution that combines Relation Siamese Net (RSN) with the clustering strategy to discover new facts. Two deep learning paradigms are illustrated in Fig. \ref{fig:figure431}.

\begin{figure*}[!t]
 \centering
 \includegraphics[width=\linewidth]{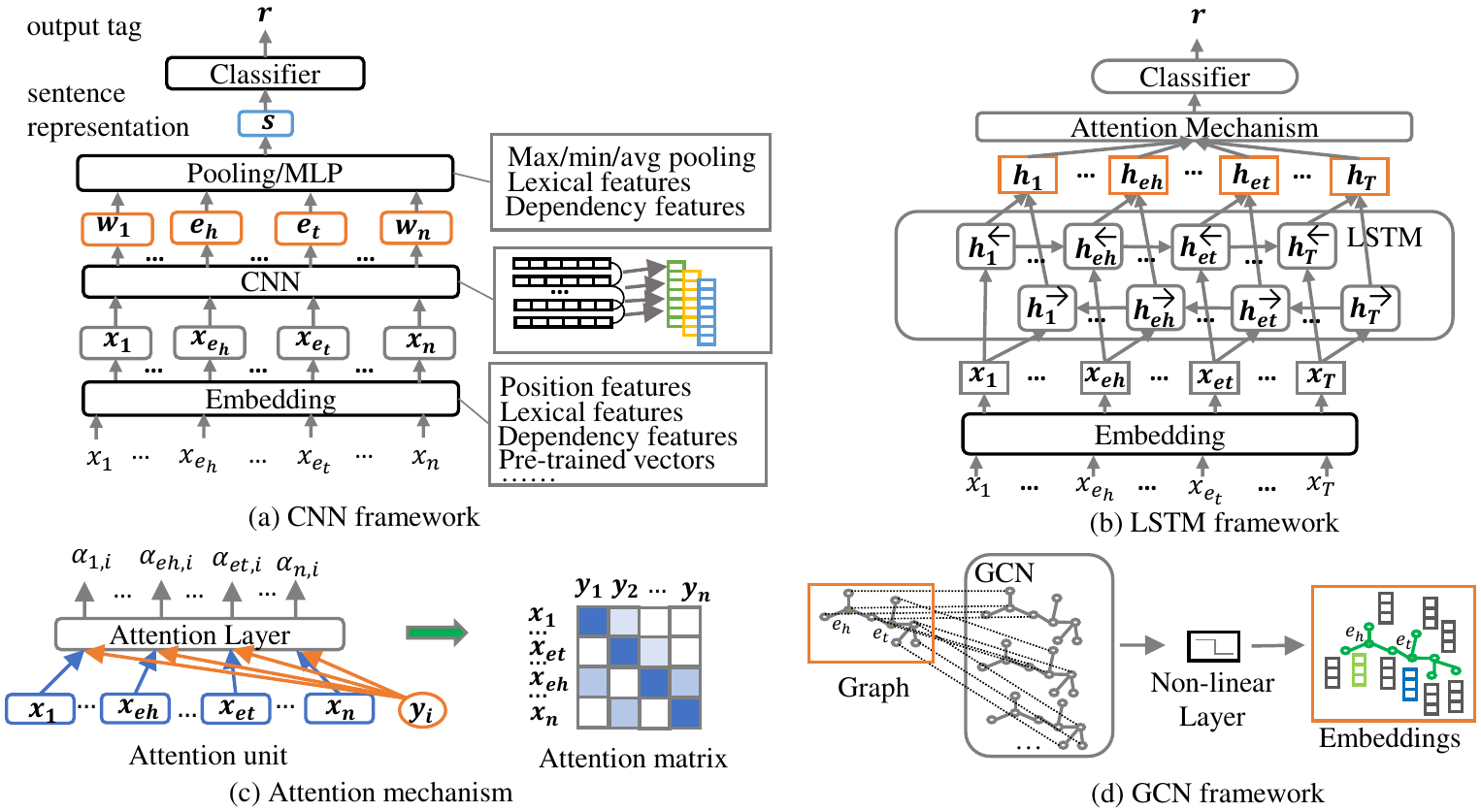}
 \caption{Some frameworks of classic relation classification models. In this figure, (a), (b), (d) display the architectures of CNNs, LSTMs, GCNs, while (c) displays the structure of the widely-used attention mechanism.}
 \label{fig:figure432}
  \vspace{-0.36cm}
  
\end{figure*} 

\begin{table}[!t]
 \centering
 
 \caption{Comparison of designs of classical and recent sentence-level RC models (arranged in terms of publication year in each category). }\label{tab:RCModelPerformance}
 
 \resizebox{\textwidth}{!}{
 \begin{tabular}{ccccc}
 \toprule[1.75pt]
 Category & Model & Architecture & Background Information\\
 \midrule[1.75pt]
 \multirow{4}{*}{Local context-aware } & CNN \cite{DBLP:conf/coling/ZengLLZZ14} & CNN + max pooling & WordNet hypernyms, position features\\
 & Multi-kernel CNN \cite{DBLP:conf/naacl/NguyenG15} & multi-kernel CNN + max pooling & Position embeddings\\
 & Attention-CNN \cite{DBLP:conf/coling/ShenH16} & CNN + word-level attention + MLP & Pre-trained word-vectors, position embeddings, WordNet, POS\\
 & Multi-level Attention \cite{DBLP:conf/acl/WangCML16} & word-level input attention + CNN + attention-based pooling & Pre-trained word-vectors, position embeddings\\
 \hline
 \multirow{2}{*}{Global context-aware models} & BiLSTM + Att \cite{DBLP:conf/acl/ZhouSTQLHX16} & BiLSTM + word-level attention & Pre-trained word-vectors\\
 & TreeLSTM \cite{DBLP:conf/acl/MiwaB16} & BiTreeLSTM + compound label embedding & SPTree, WordNet, position embeddings\\
 \hline
 \multirow{3}{*}{Graph context-aware models}
 & EPGNN \cite{DBLP:conf/acml/ZhaoWGL19} & BERT + CNN (sentence encoder)/GCN (topological encoder) & Pre-defined entity pair graph, pre-trained model, position embeddings\\
 
 & AGGCN \cite{DBLP:conf/acl/GuoZL19} & GCN + Multi-head Attention + DC + FF & pre-trained word-vectors, position features, dependency graph\\
 
 & RIFRE \cite{DBLP:journals/kbs/ZhaoXCLG21} & BERT + HGCN & pre-trained model\\
 \hline
 Task conversion-based & QA \cite{DBLP:journals/corr/abs-2010-04829} & BERT + Span Prediction Model & Converted questions/answers, pre-trained model\\ 
 \bottomrule[1.75pt]
 \end{tabular}
 }
 \vspace{-0.38cm}
\end{table}

\subsubsection{Domain-specific Relation Classification from unstructured sentence-level data} 
\
\par

 Given unstructured sentences with conceptual (entities) mentions, domain-specific relation classification tasks label the given mentions with relation tags in a pre-defined relation set given the context of the sentences. Kernel methods and deep learning frameworks typically handle relation classification as a multi-label single-class classification task.
 
SVM kernel-based methods employ the feature vectors of words to train a classifier for supervised relation classification tasks on unstructured text. These models map specific semantic objects onto a feature space via a kernel function for classification, such as with a lexical-kernel based SVM (with POS and entity tags) \cite{DBLP:conf/acl/BunescuM07}, a dependency-tree-kernel based SVM \cite{DBLP:conf/kdd/ReichartzKP10}, or a shallow-parse-tree-kernel based SVM \cite{DBLP:conf/emnlp/ZelenkoAR02}. However, a high-performance kernel function can be hard to design.

Deep learning-based frameworks automatically collect entity-related contextual information for
relation classification tasks. Given a sentence that needs its relations classified $ \{x_1, x_2, x_{eh}, ..., x_{et}, x_n\} $, let $ x_{eh}$ and $x_{et} $ stand for head and tail entities, respectively. Deep learning models will generate a representation for each word: $ \{\textbf{w}_1, ..., \textbf{e}_h, ..., \textbf{e}_t, \textbf{x}_n \}$, then the feature extractor will derive a vector $\textbf{r}$ to indicate the probability of each relation type. Models based on convolution, such as a feature-based CNN combined \cite{DBLP:conf/coling/ZengLLZZ14} with lexical features and a max-pooling strategy, focus on local contexts in neighborhood words. Nguyen and Grishman \cite{DBLP:conf/naacl/NguyenG15} use multiscale convolution windows to enhance local feature aggregation. Some studies focus on global context awareness between sentences using an LSTM framework that captures long-distance reliances. Zhou et al. \cite{DBLP:conf/acl/ZhouSTQLHX16} use a BiLSTM that employs inter-word attention to capture the long-distance dependencies of relations, while Miwa and Bansal \cite{DBLP:conf/acl/MiwaB16} incorporate tree structures into an LSTM framework. Many designs have also incorporated global context features into CNN structure via attention mechanisms to model salient interactions, such as Attention-CNN \cite{DBLP:conf/coling/ShenH16} selecting entity-relevant contexts with the word-level attention and Multi-level CNN \cite{DBLP:conf/acl/WangCML16} developing an input attention mechanism with attention-based pooling. Some of the more recent studies have explored graph-level contexts via GCNs and extracting background knowledge via pre-trained models. Examples of this approach include EPGNN \cite{DBLP:conf/acml/ZhaoWGL19}, which includes an entity pair graph for a GCN (with a pre-trained BERT model), AGGCN \cite{DBLP:conf/acl/GuoZL19}, which integrates a multi-head attention mechanism for graph convolution and RIFRE \cite{DBLP:journals/kbs/ZhaoXCLG21}, which further employs a heterogeneous graph network to merge high-order features. Cohen et al. \cite{DBLP:journals/corr/abs-2010-04829} converted relation classification into a question-answering task and incorporated BERT embeddings for classification.   Fig.
\ref{fig:figure432} depicts some of the classic frameworks of relation classification models and Table \ref{tab:RCModelPerformance} compares key design aspects of the popular models.

Some tasks require a model to handle n-ary relationships between multiple entities. To this end, semantic role labeling solutions have been devised to decompose n-ary relations into binary ones. Examples include NNF \cite{DBLP:conf/emnlp/FitzGeraldTG015} and dependency path embedding \cite{DBLP:conf/acl/RothL16}.

\subsubsection{Distant Supervised Relation Extraction/Classification}
\label{dsre}
\
\par

\begin{figure*}[!t]
 \centering
 \includegraphics[width=\linewidth]{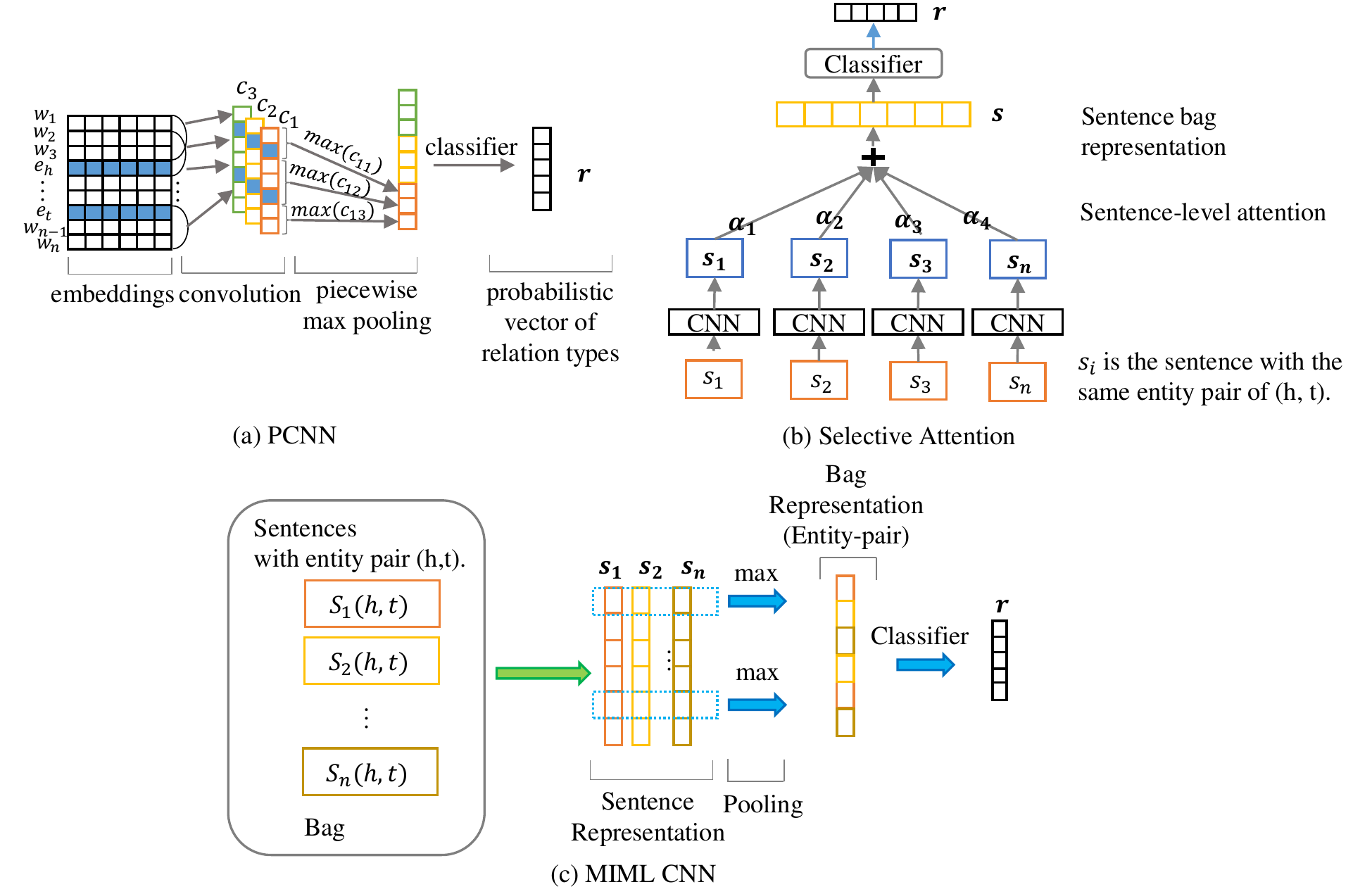}
 \caption{Milestone models for distant supervision. }
 \label{fig:figure433}
 \vspace{-0.34cm}
\end{figure*} 

Fully-supervised relation extraction/classification on large datasets will generally require a formidable amount of laborious label tagging. To cope with this problem, Mintz et al. \cite{DBLP:conf/acl/MintzBSJ09} developed a distant supervision strategy for automatically annotating relation labels with an external knowledge base (Freebase in the original work). The strategy assumes that entity pairs appearing in different sentences reflect the same relationships that link them in the knowledge base. However, distant supervision does not fully consider the data context, hence, inevitably suffers from noise. 

Data noise hampers relation classification tasks in traditional models. Some methods try to overcome this problem by enhancing the feature extractors, e.g., piecewise convolution (PCNN) \cite{DBLP:conf/emnlp/ZengLC015}. This approach divides a sentence into three separate pieces for convolution according to entities to preserve critical contextual features. See also Fig. \ref{fig:figure433}(a). Hierarchical attention mechanism \cite{DBLP:conf/emnlp/HanYLSL18} models long-tail labels to enhance decoder features. Some models involve improved learning strategies designed to promote robustness to noise. Transition matrix structure \cite{DBLP:journals/corr/LuoFWZHYZ17} learns incorrect patterns to prevent noise. Huang et al. \cite{DBLP:conf/emnlp/HuangD19} adapt collaborative learning to handle the interaction contexts, while Qin et al. \cite{DBLP:conf/acl/WangXQ18a} leverage reinforcement learning to remove wrongly-labeled samples. Further, DSGAN \cite{DBLP:conf/acl/WangXQ18} picks reliable samples for training via adversarial learning.

More recently, efforts have focused on designing instance selector structures to compare reliable features of instances in sample bags. For instance, Riedel et al. \cite{DBLP:conf/pkdd/RiedelYM10} developed the ``at-least-one'' hypothesis for multi-instance multi-label Learning (MIML). The hypothesis holds that at least one of the samples containing the same entity pairs will express the given distantly-supervised relation (i.e., the sample is correct). Based on that, selective attention \cite{DBLP:conf/acl/LinSLLS16} presents a classic design that groups sentences labeled with the same relation tags. See also Fig.  \ref{fig:figure433}(b).  MIML CNN  \cite{DBLP:conf/coling/JiangWLW16} uses a CNN to proceed with each sentence bag, then leverages a cross-sentence pooling operation to derive an entity-pair representation for multi-label relation modeling. See Fig. \ref{fig:figure433}(c). Ji et al. \cite{DBLP:conf/aaai/Ji0H017} combine entity descriptions to enhance the MIML CNN. Another direction for implementing instance-level feature extraction is instance-level attention mechanisms. The contribution of each sentence representation is then scored across different groups with the same relation tags. Last, an attention-weighted contextual representation is generated for each relation type. Many models extend this idea with MIML designs. One example is Intra/Inter-Bag Attention \cite{DBLP:conf/naacl/YeL19}. This method captures the sentence features of inner relations and outer bag-relation interactions via compound attention mechanisms and cross-relation attention \cite{DBLP:conf/aaai/YuanLTZZPWR19}, where Baye’s rule is used to acquire the global similarities of bags of different relation types.  The main goal of devising an instance selector is to emphasize instructive features in correct samples while muting dummy features in wrong-labeled data. 

Deep learning approaches also consider external knowledge to improve distance supervised relation extraction, such as incorporating the knowledge graph embeddings of entities into models \cite{DBLP:conf/naacl/ZhangDSWCZC19}. RESIDE \cite{DBLP:conf/emnlp/VashishthJPBT18} further uses a syntactic graph with side information for GCN-based representations. We compare these popular achievements in Table \ref{tab:DSREModelPerformance}.

\begin{table}[!t]
 \centering

 \caption{Comparison of popular models for distant supervision relation extraction/relation classification.}
 \resizebox{\textwidth}{!}{
 \begin{tabular}{ccccc}
 \toprule[1.75pt]
 Category & Model & Architecture & Background Information\\
 \midrule[1.75pt]
 
 \multirow{3}{*}{Enhanced feature extractor-based} 
 & PCNN \cite{DBLP:conf/emnlp/ZengLC015} & PCNN + pooling & position embeddings \\
 & TM \cite{DBLP:journals/corr/LuoFWZHYZ17} & PCNN + Transition matrix + Bag embedding & position embeddings \\
 & HAtt \cite{DBLP:conf/emnlp/HanYLSL18} & CNN/PCNN + hierarchical attention & position embeddings, relation hierarchy \\
 \hline
 \multirow{3}{*}{Enhanced learning strategy-based} 
 & RL-based \cite{DBLP:conf/acl/WangXQ18a} & RL-based data redistributor + CNN/PCNN + Result-driven reward & position embeddings \\
 & DSGAN \cite{DBLP:conf/acl/WangXQ18} & GAN + CNN/PCNN + Attention & position embeddings \\
 & CCL-CT \cite{DBLP:conf/emnlp/HuangD19} & CNN/PCNN + [Net Att + Self Att] + CCL-CT & position embeddings \\
 \hline
 \multirow{4}{*}{Instance feature-based}
 & Lin et al.\cite{DBLP:conf/acl/LinSLLS16} & CNN/PCNN + selective attention + max-pooling & position embeddings \\
 & MIML CNN \cite{DBLP:conf/coling/JiangWLW16} & CNN (sentence) + Cross-sentence max-pooling & position embeddings\\
 & Ye et al. \cite{DBLP:conf/naacl/YeL19} & CNN/PCNN + Intra/inner bag attention & pre-trained model \\
 & Yuan et al. \cite{DBLP:conf/aaai/YuanLTZZPWR19} & PCNN (Sentence) + Cross-relation Cross-bag Selective Attention & position embeddings \\
 \hline
 \multirow{3}{*}{Background information-enhanced}

 & MIML CNN + ED \cite{DBLP:conf/aaai/Ji0H017} & description embeddings + MIML CNN + & entity description, position embeddings \\
 & RESIDE \cite{DBLP:conf/emnlp/VashishthJPBT18} & Bi-GRU (sentence) + Syntactic GCNN & Dependency graph, external KB \\
 & Zhang et al. \cite{DBLP:conf/naacl/ZhangDSWCZC19} & CNN/PCNN (sentence) + GCN + Knowledge-aware attention & external KG, position embeddings \\
 
 \bottomrule[1.75pt]
 \end{tabular}
 \label{tab:DSREModelPerformance}
 }
 \vspace{-0.38cm}
\end{table}

\subsubsection{Few-shot Relation Classification}
\
\par
\label{fsrc}

Low-resource scenarios, specifically, few-shot and zero-shot relation classification, require a deep learning model to learn from a few examples. Few/Zero-shot Learning, also called meta-learning, only fuels a few samples to drive DL models, specifically, few-shot, zero-shot learning. Few-shot learning feeds a support data set in the N-way K-shot form that provides K instances for each relation type of the general N ways (N*K samples in total) and predicts data labels in the query set based on the given support set. Ulteriorly, Zero-shot learning follows the above form, but query sets contain unseen sample labels that do not appear in the support sets. It is noticeable that in the big data environment, the long-tail phenomenon exists in knowledge bases where the majority of knowledge types express with few samples. \cite{DBLP:conf/emnlp/XiongYCGW18} Meta-learning configurations commonly appear in various sub-tasks of knowledge acquisition and knowledge refinement. Generally, researchers have tried to amplify the usable characteristics of these low-resource configurations through three methods: metric-learning, meta-learning, and domain adaptation. 

  

Meta-learning enhances optimizers by reserving conveyable  meta-information from limited supervision. Model-agnostic machine learning (MAML) \cite{DBLP:conf/icml/FinnAL17} improves batch learning through a two-stage multiple gradient descent. Here, the model is trained to estimate the gradients of each relation type before all the sample types undergo general optimization with the estimations. Task-sensitive meta-information of respective relation types is hoarded in partial gradient values. Gradient estimation through separate backpropagation is also applied by other models. MetaNet \cite{DBLP:journals/corr/MunkhdalaiY17}  utilizes the fast-slow mechanism to obtain high-order implicit relational meta-features of samples and the specific task. Both the meta learner and the base learner contain a group of slow weights and fast weights for optimization. Another critical  problem is catastrophic forgetting. To deal with this issue, Wu et al. \cite{DBLP:conf/aaai/WuLLHQZX21} developed a curriculum-meta learning strategy that reviews samples in order and preserves the learned features in a memory mechanism.

\begin{figure*}[!t]
 \centering
 \includegraphics[width=\linewidth]{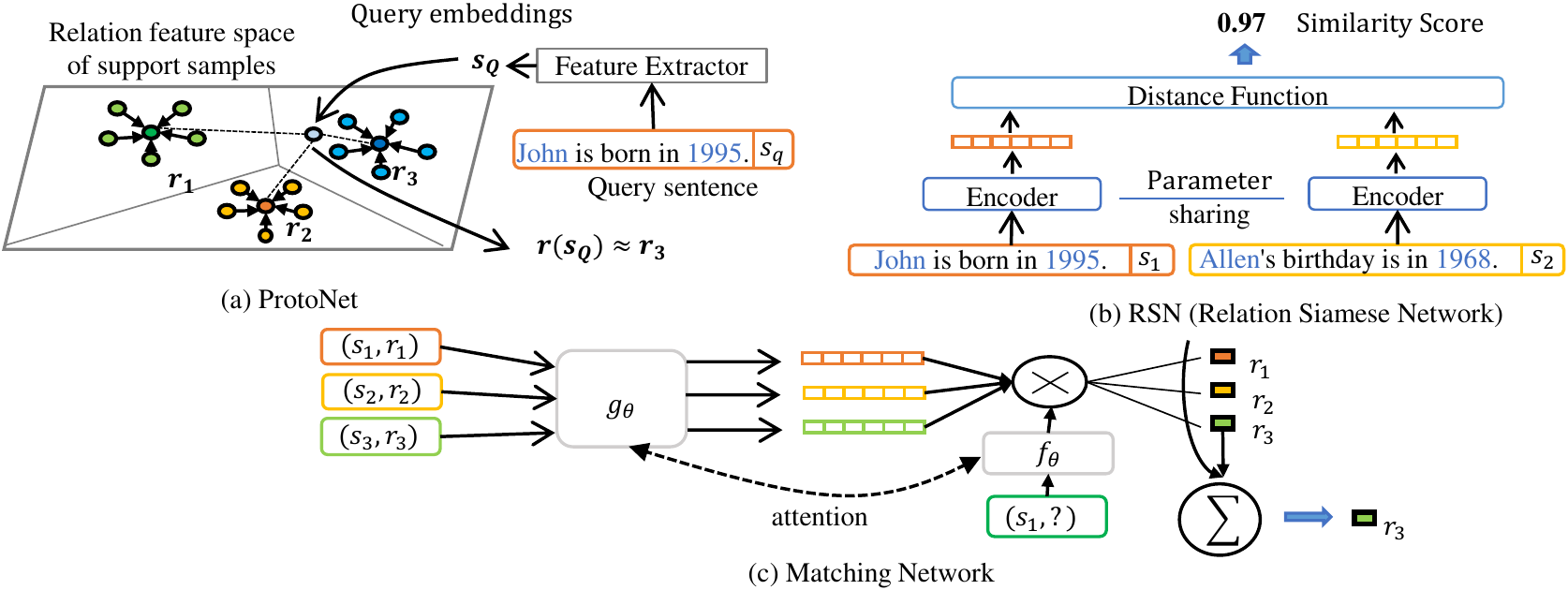}
 \caption{The metric-based few-shot relation classification models. In this figure, (a) shows ProtoNet \cite{DBLP:conf/nips/SnellSZ17}, which compares the distances of a query sample among support vectors of different relation types. (b) shows RSN \cite{DBLP:conf/aaai/GaoHX0LLS20}, which calculates the similarity of sample pairs. (c) shows Matching Network \cite{DBLP:conf/nips/VinyalsBLKW16}, which uses an attention mechanism to tag a query by matching it with different-tagged support samples.}
 \label{fig:fig4342}
 \vspace{-0.38cm}
\end{figure*}

Metric learning aims at finding metric spaces with which to compare different relation types. To determine the relation types in a query sample, ProtoNet \cite{DBLP:conf/nips/SnellSZ17}, for example, averages the embeddings of each relation type in a support set as a prototypical support vector. LM-ProtoNet \cite{DBLP:conf/cikm/FanBSL19} exploits the fine-grained features of relational context to build support vectors, concatenating phrase embeddings with sentence embeddings induced by a CNN. Noises in low-resource samples attract impertinent meta-features, reducing the robustness of deep learning models. Gao et al. \cite{DBLP:conf/aaai/GaoH0S19} combine feature-level attention with instance-level attention to emphasize reliable prototypical 
features. Furthermore, Matching Network \cite{DBLP:conf/nips/VinyalsBLKW16} presents an attention-based embedding strategy for classification, calculating the attention scores of the query sample for each support sample through vector multiplication. RSN \cite{DBLP:conf/aaai/GaoHX0LLS20} compares the similarity of the sample embeddings. Another approach, called multi-Level matching and aggregation network (MLMAN) \cite{DBLP:conf/acl/YeL19}, aggregates the local and instance features by aggregating the support vectors with the query vector to match the correct long-tail class label for the query sample. Fig. \ref{fig:fig4342} outlines the classic metric-based paradigms.

 Few-shot RC designs also consider feature augmentation strategies to mitigate data deficiency with intriguing model designs and background knowledge. Similar to \cite{DBLP:journals/corr/abs-2010-04829}, Levy et al. \cite{DBLP:conf/conll/LevySCZ17} turn zero-shot RC into a reading comprehension problem to comprehend unseen labels by a template converter. Soares et al. \cite{DBLP:conf/acl/SoaresFLK19} compose a compound relation representation for each sentence by the BERT contextualized embeddings of entity pairs and the corresponding sentence. GCNs also deliver extra graph-level features for few-shot learning. Satorras and Estrach \cite{DBLP:conf/iclr/SatorrasE18} propose a novel GCN framework to determine the relation tag of a query sample by calculating the similarity between nodes. Moreover, Qu et al. \cite{DBLP:conf/icml/QuGXT20} employ posterior distribution for prototypical vectors. Some designs also avail semi-supervised data augmentation based on metric learning. The previous Neural Snowball \cite{DBLP:conf/aaai/GaoHX0LLS20} (based on RSN) labels the query set via the Siamese network while drawing a similar sample candidate from external distant-supervised sample sets to enrich the support set.

Few-shot domain adaptation maps unseen labels for classification. BERT-PAIR 
pairs with domain adaptation strategies for unseen ``none-of-the-above'' types. Gao et al. \cite{DBLP:conf/emnlp/GaoHZLLSZ19} discuss domain adaptation for few-shot relation classification as a game process for searching domain-invariant features. They implement domain adaptation via adversarial training. More domain adaptation strategies for few-shot relation classification can be found in \cite{DBLP:conf/coling/WangHL0S18}.

\subsubsection{Joint Relation Extraction Models}
\
\par

\begin{figure*}[!t]
 \centering
 
 \includegraphics[width=\linewidth]{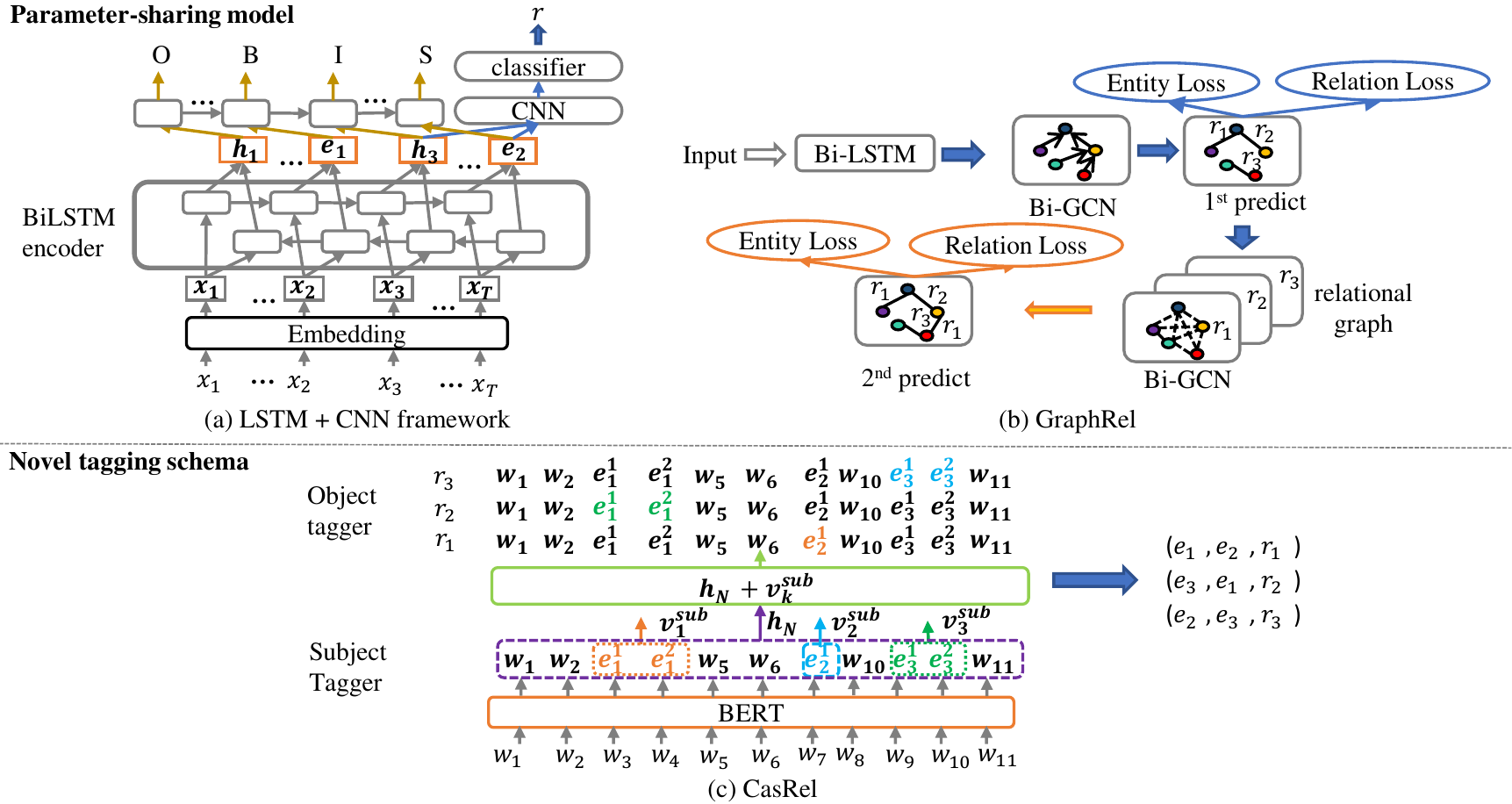}
\caption{The joint extraction model paradigms. (a) depicts \cite{DBLP:journals/ijon/ZhengHLBXHX17}, (b) depicts \cite{DBLP:conf/acl/FuLM19} and  (c) depicts \cite{DBLP:conf/acl/WeiSWTC20}. }
 \label{fig:figure4351}
\vspace{-0.38cm}
\end{figure*}

Conventional pipeline-based relation extraction (relation classification) models suffer from error propagation in each stage, while also undermining inter-task interactions. Early researchers concentrate on intriguing statistical-based features for fast end-to-end joint relation extraction, such as Integer Linear Programming-based (ILP) algorithm  \cite{DBLP:conf/conll/RothY04} solving entities and relations via conditional probabilistic model, semi-Markov chain model \cite{DBLP:conf/acl/LiJ14}  jointly decoding global-level relation features, and Markov Logic Networks (MLN) \cite{DBLP:conf/eacl/BhattacharyyaPP17} modeling joint logic rules of entity labels and relationships. Early attempts deliver prototypes of entity-relationship interactions. However, statistical patterns are not explicit for intricate contexts. The recourse for researchers has been to turn to joint extraction models, with the mainstream high-performance designs focusing on parameter-sharing strategies and novel tagging schemas.

Parameter-sharing strategies merge neural architectures for various types of tasks. They share weights and use different output layers to fetch entities that have relationships. Zheng et al. \cite{DBLP:journals/ijon/ZhengHLBXHX17} merge double BiLSTM layers for NER and RC tasks to share parameters, then use a CNN and an LSTM Network to label relationships and entities respectively. Miwa and Bansal \cite{DBLP:conf/acl/MiwaB16} also integrate the dependency features associated with NER and RC with a combination of a Bi-LSTM and a Bi-TreeLSTM layer. Some models focus on delicate strategies for distributing cross-task characteristics. The GraphRel model \cite{DBLP:conf/acl/FuLM19}, for example, intuitively leverages two-phase supervision to dedicate cross-task interaction through two respective BiGCN layers.  The GCN frameworks incorporates a dependency graph with a relation-entity graph for exploiting deep features.

To handle overlapping labels, novel tagging schemes set joint decoding targets for the output layers with compound labeling. Zheng et al. \cite{DBLP:conf/acl/ZhengWBHZX17} extend BIES labels with the relationship types and roles of a word (e.g., the subject or object of a sentence) to develop a sequence tagging task comprising named entity recognition and relation classification. Wei et al. \cite{DBLP:conf/acl/WeiSWTC20} intuitively label all object candidates of a subject entity via a cascade map function for each relation type to contain overlapping mentions. Further, Wang et al. \cite{DBLP:conf/coling/WangYZLZS20} developed a hand-shaking scheme to alleviate exposure bias within overlapping entities. Bekoulis et al. \cite{DBLP:journals/eswa/BekoulisDDD18a} devised a multi-head selection mechanism to explore all entity/relation combinations. Li et al. \cite{DBLP:conf/acl/LiYSLYCZL19} turned entity-relation tagging into a multi-turn question answering problem, leveraging the machine reading comprehensive (MRC) model for long-range semantics between entities.  Unlike previous schemes, KGGen \cite{DBLP:journals/tkde/ChenZLYJ22} directly generates triples via an encoder-decoder/generator structure based on a pre-trained model combined with adversarial learning, which overcomes feature reliance on entity co-occurrence information. Fig. \ref{fig:figure4351} shows some seminal joint extraction models.

Novel distribution embedding-based models are also proposed to model the cross-task distributions to bridge the semantic gaps between NER and RC. Ren et al. \cite{DBLP:conf/www/RenWHQVJAH17} propose a knowledge-enhanced distribution CoType model for joint extraction Task. In this model, entity pairs are firstly mapped onto their mentions in the knowledge base, then tagged with entity types and all relation candidates provided by the knowledge base. This model learns embeddings of relation mentions with contextualized lexical and syntax features while training embeddings of the entity mentions with their types, then the contextual relation mention will be derived by its head and tail entities embeddings via Translation embedding (TranE) \cite{DBLP:conf/nips/BordesUGWY13} model. The CoType model assumes interactive cooccurrence between entities and their relation labels, filling the distribution discrepancy with knowledge from the external domain and extra type features. Noticeably, this model also effectively prevents noises in distant-supervised datasets. However,  feature engineering and extra KBs are also needed.

\subsubsection{Document-level Relation Extraction}
\
\par
\label{docre}

\begin{table}[!t]
 
 \centering
 \caption{Comparison of model designs for document-level relation extraction (arranged in terms of publication year in each category). }\label{tab:DocModelPerformance}
 \resizebox{\textwidth}{!}{
 \begin{tabular}{cccccc}
 \toprule[1.75pt]
 Category & Model & Word Encoder & Long-context Encoder & Inference & Cross-sentence Feature\\
 \midrule[1.75pt]
 \multirow{4}{*}{Statistic Graph-based} & Graph LSTM\cite{DBLP:journals/tacl/PengPQTY17} & Embedding & Graph-LSTM & Softmax & Root-linked cross-sentence dependency tree\\
 & Graph-state LSTM \cite{DBLP:conf/emnlp/SongZWG18} & FFN & Graph-state LSTM & Softmax & Root-linked cross-sentence dependency tree\\
 & AGGCN \cite{DBLP:conf/acl/GuoZL19} & LSTM & GCN + Multi-head Attention & DC + softmax & Root-linked cross-sentence dependency tree\\
 & Sahu et al. \cite{DBLP:conf/acl/SahuCMA19} & embeddings & GCN & MIL-based & Coreference/Adjacent sentence edges\\

 \hline
 \multirow{8}{*}{Dynamic Graph-based Model} & GP-GNNs \cite{DBLP:conf/acl/ZhuLLFCS19} & Bi-LSTM & GCNs & MLP + softmax & Inter-node graph with generated parameters \\ 
 & EoG \cite{DBLP:conf/emnlp/ChristopoulouMA19} & BiLSTM & GCNN & Node-feature aggregation + softmax & Sentence-mention-entity pair graph\\
 & GraphRel \cite{DBLP:conf/acl/FuLM19} & BiLSTM & BiGCN & Threshold-based & Relation-weighted graph\\
 & DyGIE \cite{DBLP:conf/naacl/LuanWHSOH19} & ELmo + BiLSTM & GCN + Span enumeration & FFN & Dynamic span graph\\
 & LSR \cite{DBLP:conf/acl/NanGSL20} & BiLSTM/BERT & GCN & FFN + GCN + DC & Weighted dependency graphs\\
 & GAIN \cite{DBLP:conf/emnlp/ZengXCL20} & LSTM & GCN & FFN + Attention & hMG + EG\\
 & Xu et al. \cite{DBLP:conf/aaai/XuCZ21} & BiLSTM & AGGCN \cite{DBLP:conf/acl/GuoZL19} & LSTM + softmax & Reconstructed hetergenous S-M-E graph \cite{DBLP:conf/emnlp/ChristopoulouMA19} \\
 & DRN \cite{DBLP:conf/acl/XuCZ21} & BiLSTM & GAIN \cite{DBLP:conf/emnlp/ZengXCL20} & Aggregation + MLP & Hetergenous document-level meta-paths \\
 & RARE \cite{DBLP:conf/acl/ZhangYSXLG21} & BERT & R-GCN & MLP + softmax & Rationale graph, pre-trained model \\
 \hline
 \multirow{2}{*}{Others} 
 & ATLOP \cite{DBLP:conf/aaai/Zhou0M021} & BERT & Localized Context Pooling & Group
bilinear & Pre-train model\\
 & U-Net \cite{DBLP:conf/ijcai/ZhangCXDTCHSC21} & BERT & 2D-Conv & Matrix-based & Feature visualization\\ 
 
 \bottomrule[1.75pt]
 \end{tabular}
 }
 \vspace{-0.38cm}
 \label{Tab:docRE}
\end{table}

Entities in a document can express relationships via complex cross-sentence contexts, which defeats most of the traditional sentence-level context encoders. Novel architectures have therefore been conceived to capture document-level contexts.

Intra-sentence semantic passages are critical to document-level extraction. As such, researchers initially developed variants of an LSTM fitting graph structure to handle long-term dependencies, such as Graph LSTM \cite{DBLP:journals/tacl/PengPQTY17} and Graph-state LSTM \cite{DBLP:conf/emnlp/SongZWG18}. More recently, however, researchers have been focusing on GCN-based models to explore diverse linguistic features with novel cross-sentence graph structures.
Many approaches handle inter-sentence semantic contexts using static document graphs. For instance, for $n$-ary relation extraction, AGGCN \cite{DBLP:conf/acl/GuoZL19} links the roots of dependency trees of adjacent sentences via attention-guided GCN layers, which also overcomes the reliance on semantic role labeling. Sahu et al. \cite{DBLP:conf/acl/SahuCMA19} introduce coreference edges and adjacent word edges to form a homogeneous document graph. Christopoulou et al. \cite{DBLP:conf/emnlp/ChristopoulouMA19} employ mention/sentence/entity (M, S, E) nodes to create a heterogeneous semantic graph distinguishing various linguistic roles, while reasoning via the EoG interference layer using the above intermedia node structures.

Researchers then developed dynamic document graph models for high-order reasoning. Many models leverage dynamic edges. GP-GNNs \cite{DBLP:conf/acl/ZhuLLFCS19} deduces hidden semantic logic with dynamic edge weights in a fully-connected graph for reasoning. LSR \cite{DBLP:conf/acl/NanGSL20} regards graph structures as a latent variable to iteratively refine links and weights for constituting logical features from contexts. Xu et al. \cite{DBLP:conf/aaai/XuCZ21} considers reconstructing dependency paths to reweight relational entity pairs. GraphRel \cite{DBLP:conf/acl/FuLM19} jointly extracts entities and relationships via a two-stage procedure that incorporates static dependencies with dynamic relation-weighted graphs to enhance multi-hop reasoning. Some models also consider feature extraction with multiple graphs. For example, Zeng et al. \cite{DBLP:conf/emnlp/ZengXCL20} designed a heterogeneous mention-level graph with an entity-level graph for multi-hop inference.

Another direction for document-level relation extraction is reasoning with evidence. Zhang et al. \cite{DBLP:conf/acl/ZhangYSXLG21} develop a rationale graph with external tagged co-occurrence evidence features for capturing long-term relational dependencies. Dynamic graphs with alterable nodes have also been considered within the realms of complex reasoning. DyGIE \cite{DBLP:conf/naacl/LuanWHSOH19} prunes mute low-confident entity spans nodes through gate mechanisms for document-level feature exploitation. Other approaches seek to understand the common sense rules implicit in the different contexts. Discriminative reasoning network (DRN) \cite{DBLP:conf/acl/XuCZ21} recognizes common-sense relationships while performing intra-sentence reasoning through heterogeneous graph representation features. The method is based on the assumption that multi-scale contexts with syntactic structures contain distinguishable common-sense features. Background common-sense features can also be acquired from pre-trained models like COMET \cite{DBLP:conf/acl/BosselutRSMCC19}. However, understanding how common sense is expressed over contexts and how it unfolds in human logic remains challenging.

U-Net \cite{DBLP:conf/ijcai/ZhangCXDTCHSC21} employs a U-shaped segmentation for document-level reasoning via a multilayer convolution. Further, it treats a document as visual semantic information. ATLOP \cite{DBLP:conf/aaai/Zhou0M021} introduces localized context pooling to distill the entity-relevant features of BERTs while using an adaptive threshold for decoding reliable relations. This technique does not rely on graph structures. We compare the designs of the typical milestone models in Table \ref{Tab:docRE}.

\section{Knowledge Graph Refinement from Structured Data} \label{Section 5}

\begin{figure*}[!t]
 \centering
 \includegraphics[width=\linewidth]{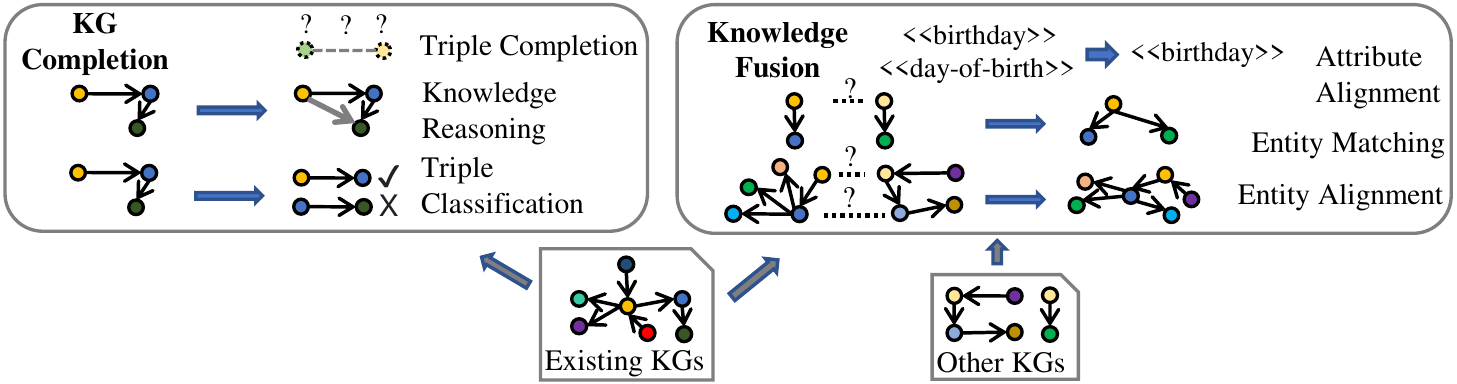}
 \caption{Illustration of knowledge refinement.}
 \label{fig:figure6}
 \vspace{-0.38cm}
\end{figure*} 

Raw knowledge graphs constructed from unstructured or semi-structured data can be sparse, and the knowledge triples can be incomplete or corrupted. Knowledge graph refinement repairs these problems through background semantics or by populating knowledge triples with additional knowledge graphs (structured data). The sub-tasks of knowledge graph refinement include knowledge graph completion and knowledge fusion. The general procedure is shown in Fig. \ref{fig:figure6}.

\subsection{Knowledge Graph Completion}
Knowledge graph completion fills in incomplete triples while deriving new triples from completed ones. In terms of completed triples, knowledge graph completion evaluates the accountability of each triple through triple classification. By accountability, we mean the correctness of the triples.
\subsubsection{Embedding-based Triple Completion}
\
\par

An embedding-based link prediction model leverages distribution representations to search for elements that can fill missing parts formulated as (h, ?, t) or (?, r, t) (entity prediction), and (h, ?, t) (relation prediction). For example, the TransE-based model \cite{DBLP:conf/cikm/Moon0S17} searches the head entity h, the tail entity t, and the relation r, whose representations approach $\textbf{h} + \textbf{r} = \textbf{t}$ to complete a triple. Later, researchers discovered that the previous symmetrical TransE model does not consider one-to-many relationships. Focus then turned to importing hyperspace structures with distance-based translation models for link prediction, such as TransR \cite{DBLP:conf/aaai/LinLSLZ15}, TransH \cite{DBLP:conf/aaai/WangZFC14}, and TranSparse \cite{DBLP:conf/aaai/JiLH016}. Some models, such as RESCAL \cite{DBLP:conf/icml/NickelTK11}, TuckerER \cite{DBLP:conf/emnlp/BalazevicAH19}, DistMult \cite{DBLP:journals/corr/YangYHGD14a} and NTN \cite{DBLP:conf/nips/SocherCMN13}, consider matching entity pair representations to a latent relational semantic space for predictions with large graphs.
 
More recently, researchers have focused on semantic knowledge structures. HAKE \cite{DBLP:conf/aaai/ZhangCZW20}, for instance, uses the polar coordinate system to model semantic hierarchies in knowledge graphs, like hypernyms, hyponyms, and the apposition of an entity’s ontological associations, which differentiate various-layered entity vectors by mold and angle constraints. CAKE \cite{DBLP:journals/corr/abs-2202-13785} boosts negative sampling with common sense rules. Many models, such as SimKGC \cite{DBLP:journals/corr/abs-2203-02167} and HaLE \cite{DBLP:conf/www/Wang0S22}, optimize negative sampling for low-dimension embeddings via contrastive learning. CAFE \cite{DBLP:journals/eaai/BorregoAHRR21} introduces a neighborhood sub-graph feature set to enhance relevant link information. Further, there has been interest in decomposing the semantic constituents of knowledge representations with the sub-structures of knowledge graphs via semantic diffusion mechanisms of GCNs. DisenKGAT \cite{DBLP:conf/cikm/WuSCCLZW021} discerns the high-order neighbor node features of a knowledge graph by disentangling the representation components into distinct semantics implied in the sub-graph structures. The hypothesis behind these models is that a large knowledge graph should contain sufficient subsets that can be reduced into k components to reason about linked entity nodes.

\subsubsection{Relation Path Reasoning}
\
\par
Relation path reasoning deduces new facts through completed triple sequences as support evidence, such as ``(B, lives-in, Seattle)← (A, works-in, Microsoft), (Microsoft, located-in, Seattle)''.

Early attempts develop random-walk models for relation path reasoning that infer relational logic paths in a latent variable logic graphic model. Path-Ranking Algorithm (PRA) \cite{DBLP:journals/ml/LaoC10} generates a feature matrix to sample potential relation paths. However, the feature sparsity in the graph impedes random walk approaches. Semantic enrichment strategies are proposed to mitigate this bottleneck, such as inducing vector space similarity \cite{DBLP:conf/emnlp/GardnerTKM14} and clustering associated relations \cite{DBLP:conf/acl/WangLLWL16}. 

Later, researchers model the relation path reasoning tasks as a Markov decision process so as to recognize logical constraints within the knowledge environment. Deep reinforcement learning achieves this idea by learning a policy agent that assesses each selection step and expands the reasoning path. DeepPath \cite{DBLP:conf/emnlp/XiongHW17} models the state space as (pre-trained) translation-based representations of entities and their induced relations. The taken actions then find the best matching relation labels via the feature space of entity pairs. Rewards for actions are calculated by a binary function. However, low-quality evaluations by the binary reward function will mean a RL-based  model that is not well generalized to handling incomplete knowledge structures \cite{DBLP:conf/emnlp/LinSX18}. To this end, Lin et al. \cite{DBLP:conf/emnlp/LinSX18} devised a soft reward shaping function based on the vector spaces of relations and entities, while Li et al. \cite{DBLP:conf/icdm/LiJGWC18} employ multiple agents to select entity pairs and relations. M-Walk \cite{DBLP:conf/nips/ShenCHGG18} leverages an RNN to capture chronological state dependencies among pathing decisions. 

 More designs leverage neural networks that capture global features to find reasonable paths. Path-RNN \cite{DBLP:conf/acl/NeelakantanRM15} recursively aggregates relation path features for multi-hop reasoning. The chains-of-reason model \cite{DBLP:conf/eacl/McCallumNDB17} enhances a path-RNN with attention mechanisms to emphasize multiple-path dependencies with type information in the entities. Chen et al. \cite{DBLP:conf/aaai/ZhangDKSS18} unify path-reasoning and path-finding tasks via variational encoding.

Some methods further focus on attention mechanisms to augment features for reinforcement learning. ADRL \cite{DBLP:journals/tpds/Kardani-Moghaddam21} leverages a self-attention mechanism to emphasize neighborhood entity-relation interaction features. Similarly, Wang et al. \cite{DBLP:conf/emnlp/WangLPM19} introduce a graph attention mechanism to enhance knowledge features. Recent research interest has been drawn into incorporating neural structures that handle intricate semantic features, such as Zheng et al.’s hierarchical policy network \cite{DBLP:conf/ksem/ZhengZC21} and DAPath \cite{DBLP:journals/nn/TiwariZP21}, which incorporates a distant-aware mechanism to issue rewards via path length features. MemoryPath \cite{DBLP:journals/ijon/LiWPM21} is an attention-based memory component that preserves knowledge features for reinforcement learning and alleviates the model’s reliance on pre-trained embeddings.

Many efforts also focus on automatically mine logic rules to pave reasoning paths. There are methods for rule discovering, such as AMIE \cite{DBLP:conf/www/GalarragaTHS13}, RLvLR \cite{DBLP:conf/ijcai/OmranWW18} and RuleN \cite{DBLP:conf/semweb/MeilickeFWRGS18}. Instead of searching for promising relation path patterns approaching the symbolic essence of knowledge, the rule mining approaches extract and prune logic rules from a reasonable KG structure, then perform link prediction via the collected rule templates. However, unseen knowledge paths cannot be easily derived by logical rules in incomplete graphs.

Another research direction is to fuel logic rules into neural models to boost path reasoning. KALE \cite{DBLP:conf/emnlp/GuoWWWG16} jointly embeds first-order logic rules with knowledge embedding to enhance relation inference. RUGE \cite{DBLP:conf/aaai/GuoWWWG18} iteratively rectifies KG embeddings via learned soft rules, then performs relation path reasoning. Logic rules are also leveraged as the side semantic information into neural models. NeuralLP \cite{DBLP:conf/nips/YangYC17} proposes a neural framework that encodes logic rule structures into vectorized embeddings with an attention mechanism. pLogicNet \cite{DBLP:conf/nips/Qu019} introduce the Markov logic network to model uncertain rules for reasoning. ExpressGNN \cite{DBLP:conf/iclr/ZhangCYRLQS20} further employs GCNN to solve neighborhood graphic semantics with logic rules. These rule-based neural models are also regarded as the application of differentiable learning availing for gradient-based optimization algorithms on logic programming.
\subsubsection{Interpretable Relation Reasoning}
\
\par
\label{interp}
Interpretability serves to make a machine learning model understandable to human users \cite{DBLP:journals/ai/Miller19}, and this plays a critical role in assessing a model’s reliability and ability to respond to different data environments. Interpretation models include self-explained pre-hoc models and inspectable post-hoc models.

Pre-hoc reasoning models that comprise transparent decision processes can be self-interpreted through their inner structures by introspection. Logic rule-mining approaches such as AMIE \cite{DBLP:conf/www/GalarragaTHS13} and RLvLR \cite{DBLP:conf/ijcai/OmranWW18} can feed back the logic rules to explain linkage decisions to users. Some models only contain some components that are interpretable to humans (e.g., the learned rules). Users can observe these learned rules as side information when reasoning with rule-finding approaches based on neural models such as NeuralLP \cite{DBLP:conf/nips/YangYC17}, pLogicNet \cite{DBLP:conf/nips/Qu019} and ExpressGNN \cite{DBLP:conf/iclr/ZhangCYRLQS20}. However, these neural networks are still black-boxed. Mainstream partial pre-hoc models also include models based on random-walk (probabilistic values for potential paths), reinforcement learning (reward values for each action), and attention (attention score for salient correlation). 

Post-hoc interpretation methods develop proxies to probe into implicit features in black-box models like matrices and neural network frameworks. Some proxies extract rules or learn a probabilistic distribution to reproduce a model. Carmona  et al. \cite{DBLP:conf/aaaiss/CarmonaR0015} train Bayesian networks with first-order logic to extract rules from embedding models. OXKBC \cite{DBLP:conf/akbc/NandwaniGACSM20} generates plausible explanation paths through the similarities between relationships and entities. Model simplification cannot decompose the features of non-linear neural models that are entwined. One solution is to conduct a sensitivity analysis to exploit the deep features. The analysis would involve imposing small perturbations on the models so as to observe how the output changes. These changes reveal the influential features. GNNExplainer \cite{DBLP:conf/nips/YingBYZL19} explores sub-graph structures that affect single-instance and multi-instance predictions. CRIAGE \cite{DBLP:conf/naacl/PezeshkpourT019} generates false facts to evaluate model performance and to locate obtrusive fact triples for each relation.

\subsubsection{Triple Classification}
\
\par
Triple classification aims to distinguish triples with surety from abnormal (untrue) triples in a knowledge graph. Many semantic models are designed for this task of judging suspicious triples in a knowledge graph that is constantly updated with novel relation types and facts.

Negative triple samples give knowledge representation models expressiveness to judge disordered triples. CKRL \cite{DBLP:conf/aaai/Xie0LL18}, for example, includes an index system for determining reliable triples, including local triple confidence, which compares the distance between a triple and a negative sample; global path confidence, which tests the global resources of the reasoning paths that form a triple; and adaptive path confidence, which scores a local reasoning path deriving a triple.

However, many potentially reasonable triples are not covered due to insufficient negative sampling – specifically, one-to-many relations \cite{DBLP:journals/corr/abs-2002-00388}. Hence, researchers have leveraged more sophisticated semantic structures to alleviate this issue. In this vein, Dong et al. \cite{DBLP:conf/aaai/DongWLBC19} expand entity embeddings into n-ball structures that are leveraged to incorporate fine-grained type chains as a way to classify triples. Amador-Domínguez et al. \cite{DBLP:journals/isci/Amador-Dominguez21} add ontological information to enhance model-agnostic expressiveness. Some models focus on advanced neural network embeddings to detect credible triples. For example, R-MeN \cite{DBLP:conf/acl/NguyenNP20} captures latent dependencies among triples by employing a multi-head attention mechanism that generates memory-based embeddings.

\subsection{Knowledge Fusion}
Real-world knowledge is usually open for updates. In most scenarios, users should be able to add external knowledge to enrich existing external knowledge graphs. In this way, knowledge fusion is designed to merge semantically-equivalent elements such as ``Trump'' and ``Donald Trump'' so as to integrate new knowledge within novel concepts or facts. The sub-tasks of knowledge fusion include attribute alignment, entity matching with small-scale incoming triples, and entity alignment with a complete knowledge graph.

\subsubsection{Attribute Alignment}
\
\par
An attribute triple indicates a property of a concept with a description value like a color, date, number, or character string. Users may use different terms to refer to the same attribute, such as ``birthday'' and ``date of birth'', where synonyms may lead to semantic sparse. Attribute alignment is thus purposed to unify attribute notations.

Many methods focus on aligning the semantic embeddings of attributes, with the premise being that two attribute names should be identical if their embeddings are close to each other. Some models leverage the similarity between attribute name strings to generate distributional embeddings, such as in \cite{DBLP:conf/ksem/SunZW20} and \cite{DBLP:conf/dasfaa/HeLQ0LZ0ZC19}. Yang et al. \cite{DBLP:conf/emnlp/YangZSLLS19} leverage a bag-of-words model to learn the contextual embeddings of attributes. Similarly, JAPE \cite{DBLP:conf/semweb/SunHL17} leverages the Skip-gram model for attribute embedding to model co-occurring attributes that are frequently used together to describe an entity, such as ``latitude'' and ``longitude'' for a position.

Attribute embeddings also provide side information for entity alignment tasks, such as definitions and descriptions. However, an attribute can also carry data that is not particularly informative, like a telephone number, which can be challenging when attempting to generate knowledge-level representations. Some models then consider using neural networks to generate embeddings based on contextual values. For example, AttrE \cite{DBLP:conf/aaai/TrisedyaQZ19} embeds each character of an attribute value with an LSTM framework so as to compose an attribute embedding for predicting potential phases in monolingual expressions. The approach incorporates an attribute-name predicate alignment strategy to handle unseen attributes. 

\subsubsection{Entity Matching with Small-scale Knowledge Graph}
\
\par
In its preliminary stages, a knowledge base will only contain a few triple mentions with insufficient information for rigorous concepts. Therefore, entity matching models integrate multi-source knowledge with the available linguistic information in small-scale data. The more recent models treat entity matching as a machine learning classification task. For example, Magellan \cite{DBLP:journals/cacm/Tan20} integrates multiple similarity functions with random forest, such that the approach also considers numerical attributes. MSejrKu \cite{DBLP:conf/semeval/SchlichtkrullA16} explores the feasibility of leveraging the classifier layer including the logic regression and MLP classifier to judge identical entity pairs. DeepMatcher \cite{DBLP:conf/sigmod/MudgalLRDPKDAR18} is a deep learning system that incorporates an RNN structure with attention mechanisms to represent attribute words for entity matching. Compared to conventional models, models based on deep learning are better at handling noise in text, especially the concept-enrichment tasks \cite{DBLP:conf/semeval/JurgensP16} with WordNet.

\begin{figure*}[!t]
 \centering
 \includegraphics{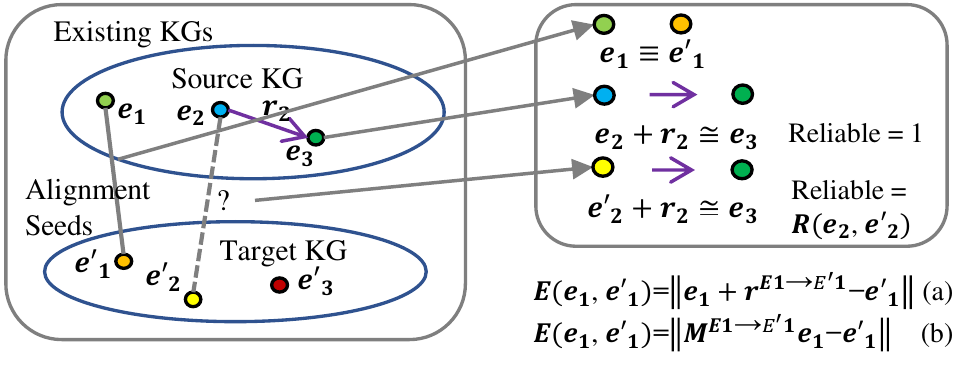}
 \caption{IPtranE \cite{DBLP:conf/ijcai/ZhuXLS17}. IPtranE scores entity pairs via (a) translation models  and (b) linear transformation models, and merges identical pairs via hard or soft alignment.}
 \label{fig:figure5231}
\vspace{-0.4cm}
\end{figure*}

 Early attempts also aim at the unique attributes of entities for entity matching. Many models leverage distance-based approaches to distributional representations of entity descriptions or definitions. VCU \cite{DBLP:conf/semeval/McInnes16} proposes first-order and second-order vector models to embed the description words of an entity pair for comprehensively measuring the conceptual distance. TALN \cite{DBLP:conf/semeval/AnkeRS16} leverages sense-based embedding derived by BabelNet to combine the definitional description of words, which first generates the embedding of each filtered definition word combing with POS-tagger, syntax feature via BabelNet, then averages them to obtain a centroid sense to obtain the best matching candidates. String-similarity-based models available for entity matching also include TF-IDF \cite{DBLP:journals/jd/Jones04}, I-Sub \cite{DBLP:conf/semweb/StoilosSK05}.

  Graph-based methods achieve feasible performance for entity matching on the small-scale KG that can consist of hierarchical graph structures. ETF \cite{DBLP:conf/sigir/VedulaNADS018} learns concept representations through semantic features and graph-based features, including Katz similarity, random walk betweenness centrality, and information propagation score.  ParGenFS \cite{DBLP:conf/fuzzIEEE/FrolovNFM19} leverages a graph-based fuzzy cluster algorithm to conceptualize a new entity. This method stimulates the thematic distribution to acquire distinctive concept clusters to search the corresponding location of an entity update in a target knowledge graph.

 Entity matching tasks can also be handled by text-similarity-based models that detect surficial similarity between entities when considering the trade-off between performance and computation cost. Rdf-ai \cite{scharffe2009rdf} proposes a systematic model to match two entity node graphs, which leverages the string-matching and lexical-feature-similarity comparing algorithms to align available attributes, then calculates the entity similarity for alignment. Similarly, Lime \cite{DBLP:conf/ijcai/NgomoA11} further leverages metric spaces to detect aligned entity pairs, which first generate entity exemplars to filter alignable candidates before similarity computation for entity fusion. Different from small-scale KGs, the shaped large KGs contain meaningful relational paths and enriched concept taxonomy. HolisticEM \cite{DBLP:conf/bigdataconf/PershinaYC15} employs IDF score to calculate the surficial similarity of entity names for seed generating and utilizes Personalized PageRank (PPR) to measure distances between entity graphs by respectively traversing their neighbor nodes.

Autonomous communities may input unique information to a KG system, such as nicknames, telephone numbers, and other personalized data. Such knowledge can only be known by users. Strategies to detect unique missing parts and ask users to fill them in are necessary.  Active learning methods \cite{DBLP:conf/ecir/BerrendorfFT21} that judge information or solve conflicts by querying users are the most reliable solutions and are indispensable in these scenarios.

\begin{figure*}[!t]
 \centering
 \includegraphics[width=\linewidth]{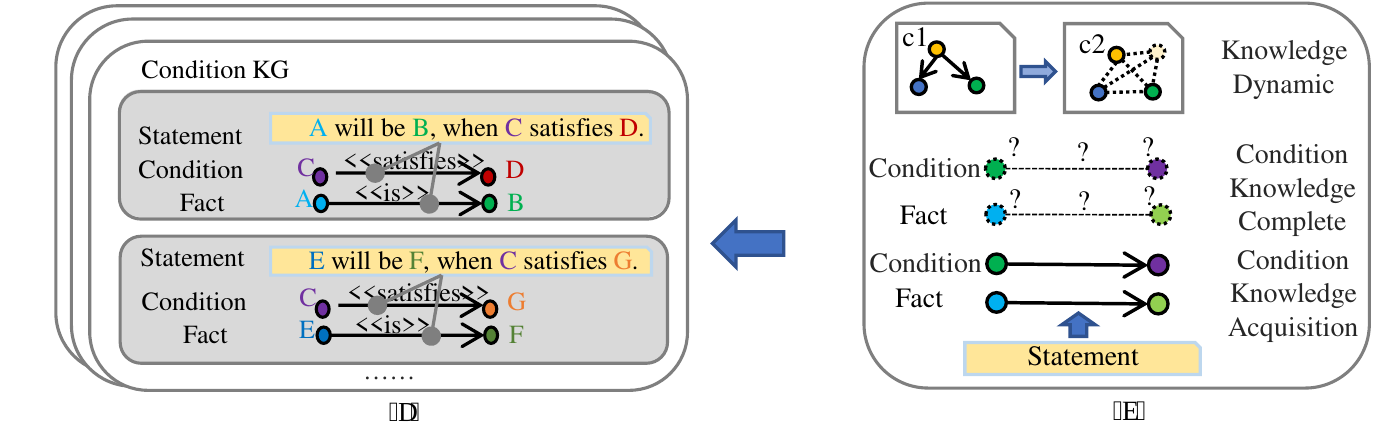}
 \caption{Knowledge evolution. Evolution analysis tasks presented in (b) manufacture data into groups of knowledge graphs (either conditional or fact knowledge graphs) displayed in (a) to portray knowledge under various dynamic conditions.}
 \label{fig:figure7}
 \vspace{-0.78cm}
\end{figure*} 

\subsubsection{Entity Alignment with Large-scale Knowledge Graph}
\
\par

Large-scale knowledge graphs usually comprise sufficient property information and graph structures that can form knowledge-aware structures with conceptual entities and relational links. Entity alignment tasks aim to integrate structured data with well-built large-scale knowledge graphs containing semantic structures at the knowledge level.

Embedding-based models learn inter-graph entity mappings for entity alignment tasks via seed entities that have the knowledge embeddings of triples. Sun et al. \cite{DBLP:conf/ijcai/SunHZQ18} point out that vanilla negative samples for link prediction can impair the ability to distinguish different entities of the same type. Hence, they use near entities in the feature space of a corresponding target entity to generate negative samples. IPTransE \cite{DBLP:conf/ijcai/ZhuXLS17} is an iterative joint embedding strategy for knowledge representation and learning entity mappings. It leverages a path translation embedding approach to embed different relation paths linking the same entity pair. These are regarded as links with identical effects. A soft alignment strategy is then used to alleviate matching errors. See Fig. \ref{fig:figure5231}. MultiEA \cite{DBLP:conf/ijcai/ZhangSHCGQ19} considers the multi-view features of entity graph attributes, links, and neighbor nodes. BootEA \cite{DBLP:conf/ijcai/SunHZQ18} includes a bootstrapped ``likely alignment'' labeling algorithm that iteratively adds reliable seeds for aligning. In cross-lingual scenarios, MtransE \cite{DBLP:conf/ijcai/ChenTYZ17} generates axis calibration and translation vectors to model feature space invariance in different languages. Additionally, some models consider self-supervision strategies to exploit seed information, such as SS-AGA \cite{DBLP:journals/corr/abs-2203-14987} and SelfKG \cite{DBLP:conf/www/LiuHWCKD022}.

One critical challenge with entity alignment is that many entities do not possess surface or structural distribution features. Thus, many entity alignment models also use attribute representation to augment the features. KDCoE \cite{DBLP:conf/ijcai/ChenTCSZ18}, for example, leverages a co-training strategy with description attributes. JarKA \cite{DBLP:conf/pakdd/ChenZT0L20} models interactions among attributes in a sparse multi-lingual knowledge graph to infer equivalent entities. Some models leverage deep learning-based neural networks for attribute context embeddings. For example, AttrE \cite{DBLP:conf/aaai/TrisedyaQZ19} leverages an LSTM to derive the dependency features of attribute values. Unlike previous methods, JAPE \cite{DBLP:conf/semweb/SunHL17} consolidates attribute embeddings with overlay relationship graph structures to capture cross-lingual disparities.

Another challenging issue is semantic graph structures for alignment. GCN-Align \cite{DBLP:conf/emnlp/WangLLZ18} was the first to propose a GCN-based framework for entity alignment tasks. Since then, recent research has focused on complicated graph semantics using GCN-based models. For instance, RNM \cite{DBLP:conf/aaai/00020WD21} matches neighborhood nodes features to compare entity pairs. RDGCN \cite{DBLP:conf/ijcai/WuLF0Y019} leverages a dual relation graph to solve contradictory representations in triangular entity graph structures.

Large-scale knowledge graphs typically contain distinctive semantic sub-graph structures for alignment. Here, graph matching neural network (GMNN) \cite{DBLP:conf/acl/XuWYFSWY19} builds a topic entity graph that links neighboring nodes to merge identical entities. AttrGNN \cite{DBLP:conf/emnlp/LiuCPLC20} partitions a knowledge graph according to attribute triple types to understand heterogeneous entity information.

Recent research direction also aims at modeling cross-graph interaction. MuGNN \cite{DBLP:conf/acl/CaoLLLLC19}, for example, proposes a cross-knowledge graph attention mechanism with a multi-channel GNN encoder that can model inter-graph structural features consistently. Similarly, GTEA \cite{DBLP:conf/IEEEwisa/JiangNSKY21} involves a joint graph attention mechanism to fuse cross-graph relational 
information.

\section{Knowledge Evolution} \label{Section 7}
Recently, researchers have focused on how knowledge evolves given environmental conditions.
Conditional knowledge graphs serve this goal by reflecting facts established under certain conditions. A conditional tuple is formulated as (h, r, t, $\gamma$), where $\gamma$ can be a prerequisite triple of a fact. Many researchers have studied this in its simplified case, as a temporal knowledge graph, where $\gamma$ is some kind of temporal information (like a timestamp) - for example {(Biden, job, vice president, 2009-2017), (Biden, job, president, 2020-)}.  Fig. \ref{fig:figure7} shows a schematic of knowledge evolution.

\subsection{Condition Knowledge Acquisition}
\label{cka}
 \begin{figure*}[!t]
 \centering
 \includegraphics{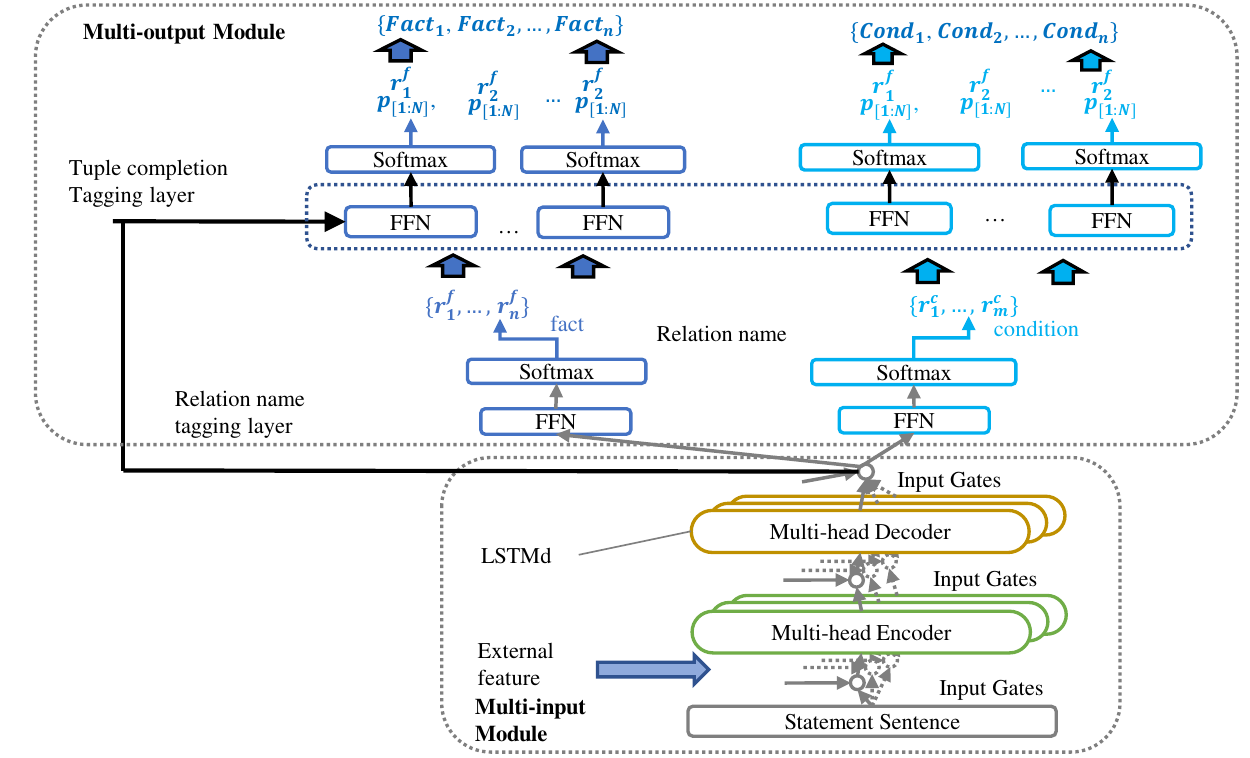}
 \caption{The architecture of MIMO model \cite{DBLP:conf/emnlp/JiangZQLCJ19} for extracting facts with conditions over texts. }
 \label{fig:figure611}
 \vspace{-0.38cm}
\end{figure*} 
Many scientific facts are established upon certain conditions, especially in the biomedical field. Early efforts have not comprehensively considered this scenario in a systematic view. Hence, Jiang et al. \cite{DBLP:conf/kdd/JiangZ00C019} developed a new tagging schema to describe conditional tuples formatted as ``B/I-XYZ'', where ``BI'' stands for positional information (begin/intermediate), ``X'' is the logic role (fact/condition), ``Y'' marks the tuple role (subject/object), and ``Z'' denotes the constituent type (concept/attribute/predicate).
Conditional knowledge extraction achieves three goals: it extracts fact tuples, it collects conditional tuples, and it connects fact conditions. Jiang et al. \cite{DBLP:conf/emnlp/JiangZQLCJ19} noted that the traditional extraction systems merge conditional information into entities to form factual triples, which will compromise entity linking. Further, the same tokens can be both subjects and objects of different tuples in an unstructured statement. They therefore devised a joint extraction method based on the multi-input multi-output sequence labeling (MIMO) to tackle this problem. Their MIMO model leverages a relation name tagging layer that denotes the relationship tags for each token via factual and conditional tagging sub-layers, respectively. A tuple completion tagging layer is then used to distinguish the logic roles of each token with different relationship names. However, Zheng et al. \cite{DBLP:conf/cicai/ZhengXLZWY21} point out that the MIMO tagging schema cannot effectively handle overlapping triples. They therefore leveraged hierarchical parsing to simplify the multi-output schema in MIMO models into a one-output schema. Fig. \ref{fig:figure611} illustrates the MIMO model.

Another popular trend in conditional knowledge extraction is temporal knowledge extraction, where a conditional triple is simplified into time. Many previous models leverage RNN structures to capture temporal dependencies and therefore identify the temporal relationships within sentences, such as \cite{DBLP:conf/acl/ChengM17} and \cite{DBLP:conf/emnlp/MengRR17}. In terms of extracting fine-grained temporal knowledge, Vashishtha et al. \cite{DBLP:conf/acl/VashishthaDW19} model events, states, and durations to match their timeline via multiple stacked attention layers. Recent research has improved solutions to handle document-level temporal knowledge extraction. For instance, TIMERS \cite{DBLP:conf/acl/MathurJDMTM20} is a rhetoric-aware graph for GCN models to interpret an intricate contiguous ``elementary discourse unit''
through the document’s expressions. Here, an elementary discourse unit is the minimal semantic unit involved in temporal activities.

\subsection{Condition Knowledge Graph Completion}
Condition Knowledge Graph Completion tasks fills incomplete triples in a Condition KG, such as (h, ?, t, $\gamma$), (h, r, ?, $\gamma$), and (h, r, t, ?).  Note that, in this section, our main focus is on methods for completing temporal knowledge graphs.

Researchers can predict incomplete temporal tuples by temporal information embedding models. TTransE \cite{DBLP:conf/www/LeblayC18} extends TransE with temporal embedding vectors. HyTE \cite{DBLP:conf/emnlp/DasguptaRT18} treats the timestamp as a hyperplane for matching entity and relation embeddings. Another promising direction is temporal-aware embeddings. In this stream, the LSTM-based model \cite{DBLP:conf/emnlp/Garcia-DuranDN18} interprets time-encoding sequences, while the CNN-based model \cite{DBLP:conf/wise/LiuHXZ19} captures the temporal consistency of contexts.

Temporal knowledge graph representations can be regarded as tensor structures along the temporal dimension, which means tensor decomposition can be used to complete temporal knowledge graphs. The main solutions for tensor decomposition include canonical polyadic decomposition and Tucker decomposition. Canonical polyadic decomposition uses the sum of several one-rank tensors to approach a target tensor. Many temporal knowledge graph completion models use canonical polyadic decomposition, e.g., \cite{DBLP:conf/ictai/LinS20} and \cite{DBLP:conf/iclr/LacroixOU20}. Tucker decomposition factorizes a target tensor into the multiplication between a kernel tensor and multiple tensors along each dimension of the target tensor. Shao et al. \cite{DBLP:journals/kbs/ShaoZYTCL22} developed a model based on Tucker-decomposition to interpret temporal semantic associations that increases the flexibility of representations that include timestamps. SpliMe \cite{DBLP:conf/ictai/RadstokCV21} obtains time-viewed entity embeddings via a static model.

Another critical topic for temporal knowledge graph completion is temporal knowledge reasoning. Recent research interest has focused on GCN-based methods. Here, Han et al. \cite{DBLP:conf/iclr/HanCMT21} exploit historical contexts by expanding a query-dependent interference subgraph based on edge attention scores. Jung et al. \cite{DBLP:conf/kdd/JungJK21} achieve multi-hop temporal reasoning via edge-based attention propagation, while Liu et al. \cite{DBLP:journals/kbs/LiuZCGXS22} enhance temporal knowledge graph reasoning via a model based on reinforcement learning. Moreover, facts in a timeline cannot ignore temporal dependencies, such as ``born-in'' before ``works-at''. Jiang et al. \cite{DBLP:conf/emnlp/JiangLGSLCS16} defines a scoring function that contains an asymmetric matrix to preserve temporal ordering constraints for reasoning.

Filling in incomplete general conditional tuples is open for further exploration. Tuples may contain more than one condition, such as chemical reactions that only occur within a certain temperature range. A systematic solution should be put into these complex scenarios. We suggest that readers also  consider causality discovery methods \cite{DBLP:journals/csur/YuGLLWLW20}.

\subsection{Knowledge Dynamic}
Many researchers have contributed to the literature on knowledge dynamics. A good proportion uses RNN structures to understand diachronic dependencies so as to predict state changes. For example, Know-evolve \cite{DBLP:conf/icml/TrivediDWS17} involves a multivariate temporal point process with an enhanced RNN structure that learns a temporal evolutionary representation function. RE-NET \cite{DBLP:journals/corr/abs-1904-05530} incorporates a neighborhood aggregator to seize concurrent interactions between entity nodes. Models have also been designed that contain evolutionary representations, such as MGraph \cite{DBLP:journals/el/LiLWLWQ21} and DyERNIE \cite{DBLP:conf/emnlp/HanCMT20}. Gracious et al. \cite{DBLP:conf/aaai/GraciousGKCD21} systematically construct a neural latent space model that combines the evolutionary information of a heterogeneous knowledge graph. Yan et al. \cite{DBLP:conf/aaai/YanLBJT21} improves a GCN model’s ability to capture topology-invariant features. The idea is to align nodes in different temporal knowledge graph snapshots and build a dynamic profile of concepts.

How knowledge evolves when different kinds of conditions change remains challenging – take the conditions needed to end the COVID-19 outbreak as an example. We recommend that readers refer to causality feature selection methods \cite{DBLP:journals/csur/YuGLLWLW20} along with experts and multi-source evidence. 

\begin{figure*}[!t]
  
  \centering
  \includegraphics[width=\linewidth]{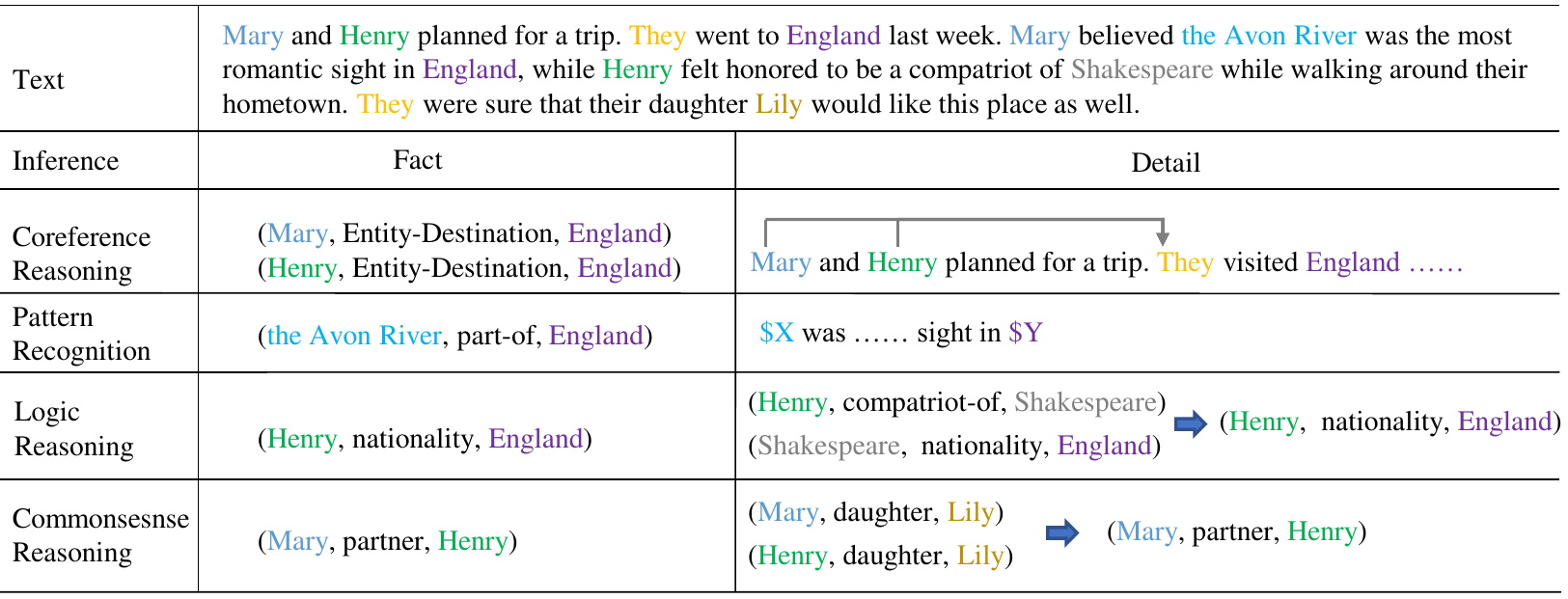}
  \caption{An example of relation inference over long context in a document.}
  
  \vspace{-0.38cm}
  \label{fig:figlcontext}
\end{figure*} 

\section{Knowledge Graph Storage} \label{Section KGStorage}
In this section, we provide a brief overview of KG storage tools for different data environments. 

Early efforts at graph storage used relational models to perpetuate constructed knowledge graphs. Traditional RDBMS provides reliable and swift CRUD operations for table-formed databases. Developers have also employed graph algorithms like depth-first traverse and shortest-path search to enhance relational databases \cite{DBLP:journals/fcsc/YanWCGZ18}. Representative examples of this type of algorithm include PostgreSQL \cite{momjian2001postgresql} and  filament \footnote{https://filament.sourceforge.net}. However, it can be very costly for a relational database to handle sparse KGs or perform data partition for distribution storage.

  Key/value databases are lightweight solutions for saving clusters in large knowledge graphs. Further, they support distributed storage with a simplified and flexible data format. Trinity \cite{DBLP:conf/sigmod/ShaoWL13} provides a high-performance in-memory Key/Value storage system to manage large knowledge graphs with billions of nodes, such as Probase. CouchDB \cite{anderson2010couchdb} uses a replication mechanism to maintain dynamic knowledge graphs. MapReduce technology automatically transforms data groups into key/value mappings. Hadoop \footnote{http://hadoop.apache.org} enables high-throughput parallel computing for knowledge graph storage via MapReduce. Pregel \cite{DBLP:journals/pvldb/ZouMCOZ11} develops a superstep mechanism to share messages between vertices for parallel computing.

   Another promising direction is to design graph databases that fit in knowledge triple structures. Neo4j\cite{DBLP:conf/oopsla/Webber12} is a lightweight NoSQL-based graph database that supports embedded dynamic knowledge graph storage. SOnes \footnote{http://github.com/sones/sones} provides object-oriented queries for KG database.  Novel languages have also been developed for knowledge storage, such as resource description framework (RDF) and web ontology language (OWL) \footnote{RDF and OWL are both standards of w3c, see also http://www.w3.org/RDF}. Some graph databases based on RDF optimize the storage of graph structures. For example, gStore \cite{DBLP:journals/pvldb/ZouMCOZ11} improves RDF-structured knowledge graph databases via sub-graph matching algorithms.

\section{Discussion on Knowledge Graph Construction} \label{Section 8}
Researchers have contributed various solutions to different aspects of knowledge graph construction. However, some challenging issues and research directions are still open for further discussion.

\begin{figure*}[!t]
 \centering
 \includegraphics{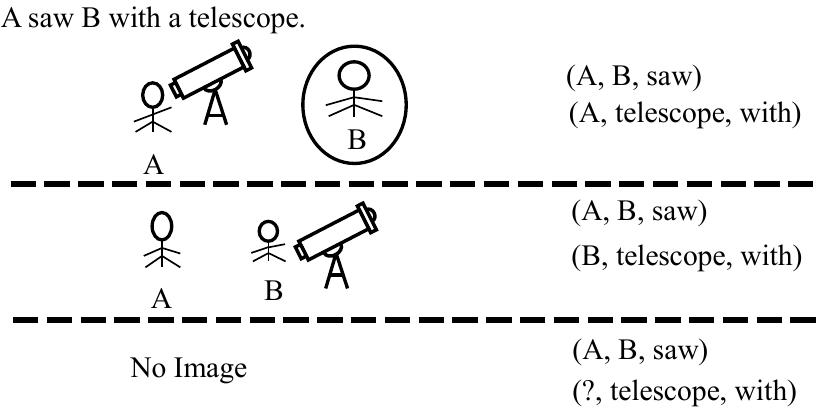}
 \caption{An example of solving information incompleteness via cross-modal dependency. In this case, the textual expression ``A saw B with a telescope.'' is unclear. An extractor can only determine the relationship between people and the ``telescope'' by interpreting the side information in the image data.}
 \label{fig:figure841}
 \vspace{-0.38cm}
\end{figure*} 

 \subsection{Long and Intricate Contexts for KG Construction}
Intricate cross-sentence or cross-paragraph contexts impedes different KG construction sub-tasks for practical use, especially relation extraction tasks. It is worth reminding readers that complex contexts do not merely relate to long-term dependency. Yao et al. \cite{DBLP:conf/acl/YaoYLHLLLHZS19} point out that four kinds of inferences include pattern recognition, coreference reasoning, logic reasoning, and commonsense reasoning, are critical to contain high-order contextual semantics. A specific example is presented in  Fig. \ref{fig:figlcontext}.

A model that handles complex long contexts should focus on intricate cross-sentence patterns while performing reasoning over multiple linguistic objects. Besides document-level extraction models in section \ref{docre}, Some efforts in section \ref{etlcontext} also model document-level contexts via heterogeneous models for entity typing. Noticeably, ambiguous expressions may occur in user-generated texts, which are usually not correctly interpreted by models without external information. Another challenging issue for reasoning is multi-hop reasoning. More linguistic structures should be explored to comprehend tortuous expressions.

Out-of-context expressions requiring background knowledge to handle are bottlenecks for KG construction. The obstacles are mainly two-fold: 1) spontaneous knowledge, and 2) evidence support.  Commonsense knowledge spontaneously generated is often utilized to derive new facts, e.g., man and woman who have kids should be couples/partners, despite such convictions sometimes inaccuracy. How to obtain commonsense rules and adapt them to suitable scenarios is an important direction. Meanwhile, many document-level datasets do not contain evidence information for correct logic paths. Efforts like \cite{DBLP:journals/corr/abs-2106-08657} have probed into document-level evidence structures for relation mentions. However, it is not likely to foresee that a model can learn to organize clues correctly to resolve facts in all scenarios (e.g., validating the conclusion in a philosophical book). We believe long-context is not merely an NLP question, and models \cite{DBLP:journals/corr/abs-2101-08091}  understanding linguistic expressions will be a critical direction.  Furthermore, conditions like temporal and geographical information in provided data sources should also be considered for rigorously comprehending contexts.

 \subsection{Multi-modal Knowledge Graph Construction and Completion}
\label{crossmodal}
Multi-modal knowledge graphs can entirely express and store heterogeneous information for display. Multi-modal knowledge can also be applied to detect fake or low-quality content, such as text with mismatched images (e.g., a document labeled Paris with a picture of London). MMKG \cite{DBLP:conf/esws/LiuLGNOR19} is a model for completing multi-modal tuples that reasons over image information, while Dost et al. \cite{DBLP:journals/dke/DostSRBS22} probe into cross-modal entity linking with text and images. Another problem for multi-modal knowledge graphs is solving semantic incompleteness in one-modal expression via 
cross-modal dependencies. We illustrate this challenge with a specific example in Fig.~\ref{fig:figure841}.

\begin{figure*}[!t]
  
  \centering
  \includegraphics{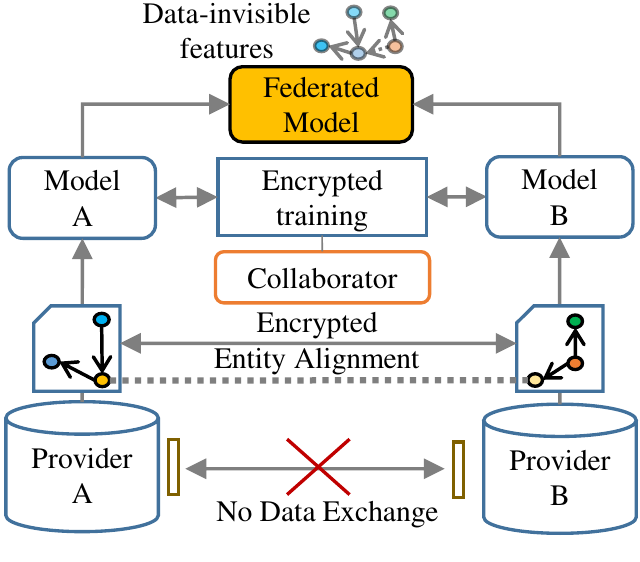}
  \caption{An illustration of building a federated model from different knowledge providers while protecting privacy. In this procedure, an encrypted entity alignment process is performed before training separate models on multi-source data parts, then a collaborator calculates and aggregates encrypted gradients of each model to prevent leakage. A federated model only reserves data-invisible crowd-sourced knowledge features.}
  \label{fig:figure852}
  \vspace{-0.38cm}
\end{figure*} 

\subsection{Federated Learning}

 Federated Learning is an enlightening direction for the essential requirements of privacy protection. A federated setting for KGs that trains model ensembles from multi-sources is one of the popular strategies. Significant advances have been conducted to federated knowledge embeddings, such as  FKGE \cite{DBLP:conf/cikm/PengLSZ021} and  FedE \cite{DBLP:conf/jist/ChenZYJC21}, which prohibit data exchange while incorporating cross-modal features during training. However, entity alignment is a paradoxical bottleneck that impedes federated learning, requiring multi-source KGs to be shared before model learning, which will exchange sensitive information during knowledge fusion. How to create a privacy-reserved super feature space for encrypted entity alignment while federating features is still open for exploration. Designing more privacy-friendly models for constructing KGs is critical for sensitive data scenarios. We illustrate the procedure of developing a federated model in Fig. \ref{fig:figure852}.

\subsection{Advanced Semantic and Dynamic in Knowledge Graph Construction Tasks}
Recent research has extended to advanced semantic evaluation tasks, such as detecting equivoque \cite{DBLP:conf/semeval/MillerHG17} and validating facts with common sense \cite{DBLP:conf/semeval/WangLJWZZ20} to handle complex lingual phenomena. Interpreting literary expressions, such as similes and metaphors, is a future direction for intelligent knowledge graph construction, e.g., ``Tom went to heaven in 2008.'' means `` Tom, died-in, 2008''. Developing or fine-tuning pre-trained models with advanced semantics will be a starting point for high productivity.

Furthermore, many studies have been conducted on the dynamics of temporal knowledge graphs. However, how knowledge semantics evolves with the general associated conditions remains an unexplored field. Diving into heuristic questions such as ``How do the professional social networks of medical staff change with the phases of a pandemic?'' may help us detect implicit factors for boosting policy making in public health. Capturing the dynamics of how associated conditions affect related facts is the ultimate direction for simulating general human knowledge.

\subsection{Human-machine Synergy for  Knowledge Graph Construction}
\label{AutonomousKGC}

\begin{figure*}[!t]
 \centering
 \includegraphics[width=\linewidth]{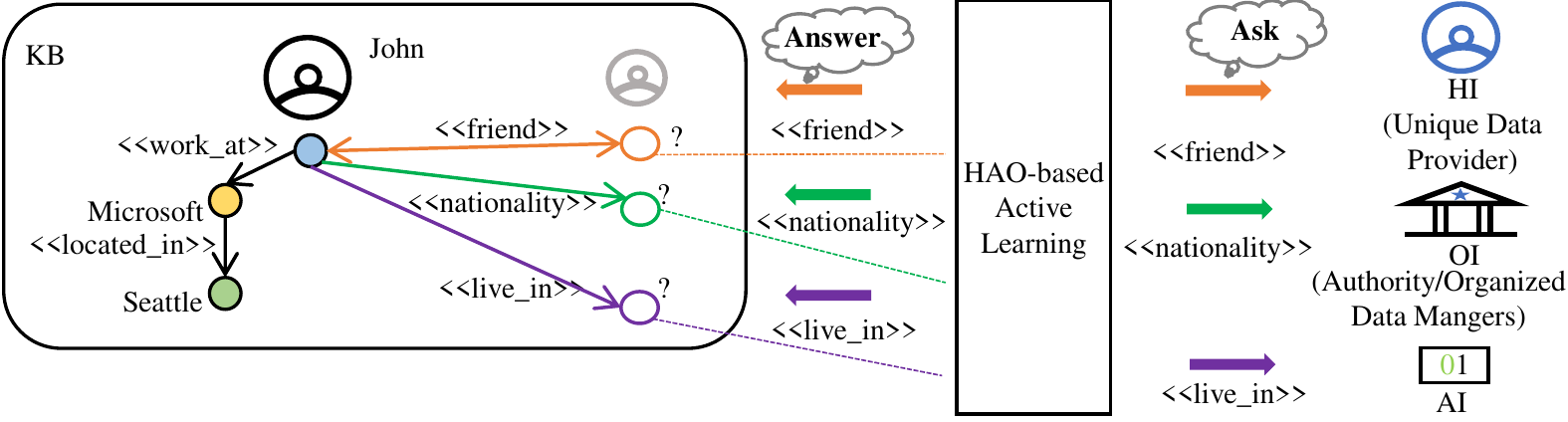}
 \caption{An HAO-based active learning case for knowledge graph construction. HAO-based active learning models select users with appropriate roles to label uncertain samples. In this case (if the privacy policy allows), John`s nationality will be labeled by the authority (OI), while his friends will be found by asking users (HI) in his social network. The AI will then derive his place of living from known facts. }
 \label{fig:figure851}
 \vspace{-0.38cm}
\end{figure*} 

Asking appropriate users to complete and correct knowledge graphs is the ultimate solution for obtaining unknown facts in the open world. To this end, Wu et al. \cite{DBLP:journals/kais/WuW19} devised the HAO model to solve different construction problems by having humans and machines collaborate. An HAO-based active learning model that automatically identifies different roles (e.g., field experts (HI), organized authorities (OI), computing systems (AI), etc.) and assigns undetermined data to appropriate users 
to tag will be a promising direction to endow wisdom to knowledge graph construction frameworks. We present an illustration of this significant idea in Fig. \ref{fig:figure851}.

\subsection{Cross-lingual Knowledge Graph}
Building cross-lingual knowledge graphs is a long-term goal that refers to integrating imbalanced resources distributed in different languages. Xlore \cite{DBLP:conf/semweb/WangLWLLZSLZT13} provides an enlightening example of aligning cross-lingual entities via deep learning approaches. However, machine translation remains a formidable bottleneck to cross-lingual tasks. Firstly, errors and conflicts generated in the process of translation will compromise the effort of refinement. Secondly, data resources expressed in minority languages may be insufficient for machine learning. To accurately perform automatic low-resource knowledge translation while resolving cross-lingual conflicts is a promising direction.

\subsection{End-to-end Unified Framework for Construction}

End-to-end extraction methods, such as GCN-based frameworks, unify the sub-tasks of knowledge acquisition into one unified extraction task, surpassing pipeline designs. However, incorporating knowledge acquisition with knowledge refinement tasks to build an integrated joint model remains a formidable bottleneck. Searching for end-to-end frameworks that unify both extraction and the refinement of knowledge graphs could be an enlightening future direction. Providing a high-quality off-the-shelf solution avoids the need for manual adjustments to components, and one that considers cross-task semantics would be a worthwhile undertaking. Further, training a framework that unifies the general procedures of knowledge graph construction would be a worth-to-solve challenging multi-task learning problem.

\section{Conclusion} \label{Section 9}

With this paper, we delivered a comprehensive survey on the topic of knowledge graph  construction. Specifically, we reviewed the tasks, methods, challenges, and related resources used to construct, refine, and integrate KGs from various data types in different scenarios. To probe into the essential topics for the big data environment, we systematically presented the paragon models for obtaining fine-grained concepts (entity typing), dealing with low-resource knowledge (extraction tasks in few-shot scenarios), understanding large linguistic objects (document-level relation extraction), complex reasoning (logic and interpretable reasoning) and handling conditional structures (temporal and general conditions) in knowledge graphs. Moreover, we provided briefs on practical KG toolkits and projects. In conclusion, knowledge graph construction has become a critical topic for enabling human intelligence in AI applications. In the future, the research community will certainly be searching for more paradigms to empower KGs with wisdom in massive heterogeneous, autonomous, complex, and evolving data environments while enhancing collaborations between knowledge communities.

\footnotesize
\bibliographystyle{ieeetr}


\end{document}